\documentclass[sigconf]{acmart}
\AtBeginDocument{%
  }

\setcopyright{acmlicensed}
\copyrightyear{2026}
\acmYear{2026}
\setcopyright{cc}
\setcctype{by}
\acmConference[CCS '26]{Proceedings of the 2026 ACM SIGSAC Conference on Computer and Communications Security}{November 15--19, 2026}{The Hague, Netherlands}
\acmBooktitle{Proceedings of the 2026 ACM SIGSAC Conference on Computer and Communications Security (CCS '26), November 15--19, 2026, The Hague, Netherlands}
\acmDOI{10.1145/3830454.3832609}
\acmISBN{979-8-4007-2871-6/2026/11}

\usepackage{tikz}
\usepackage{amsmath}

\usepackage{filecontents}

\usepackage{algorithmic}
\usepackage{amsmath,amsfonts}
\usepackage{booktabs}
\usepackage{fancyhdr}
\usepackage{float}
\usepackage{flushend}
\usepackage{graphicx}
\usepackage{mathtools}
\usepackage{microtype}
\usepackage{subcaption}
\usepackage{syntax}
\usepackage{tcolorbox}
\usepackage{textcomp}
\usepackage{url}
\usepackage{xcolor}
\usepackage{flexisym}
\usepackage{forest}

\usepackage{enumitem}
\usepackage{multirow}
\usepackage[table,xcdraw]{xcolor}
\usepackage{listings}
\usepackage{upquote}
\usepackage{adjustbox}
\usepackage{tabularray}
\usepackage{nicematrix}

\usepackage[ruled,linesnumbered,vlined]{algorithm2e}
\usepackage{mdframed}
\usepackage{pifont}

\definecolor{delim}{RGB}{20,105,176}
\definecolor{numb}{RGB}{106, 109, 32}
\definecolor{string}{rgb}{0.64,0.08,0.08}

\usepackage{color}
\definecolor{lightgray}{rgb}{0.95, 0.95, 0.95}
\definecolor{darkgray}{rgb}{0.4, 0.4, 0.4}
\definecolor{editorGray}{rgb}{0.95, 0.95, 0.95}
\definecolor{editorOcher}{rgb}{1, 0.5, 0} 
\definecolor{editorGreen}{rgb}{0, 0.5, 0} 
\definecolor{orange}{rgb}{1,0.45,0.13}		
\definecolor{olive}{rgb}{0.17,0.59,0.20}
\definecolor{brown}{rgb}{0.69,0.31,0.31}
\definecolor{purple}{rgb}{0.38,0.18,0.81}
\definecolor{lightblue}{rgb}{0.1,0.57,0.7}
\definecolor{lightred}{rgb}{1,0.4,0.5}

\usetikzlibrary{arrows.meta}

\lstdefinelanguage{CSS}{
  keywords={color,background-image:,margin,padding,font,weight,display,position,top,left,right,bottom,list,style,border,size,white,space,min,width, transition:, transform:, transition-property, transition-duration, transition-timing-function},	
  sensitive=true,
  morecomment=[l]{//},
  morecomment=[s]{/*}{*/},
  morestring=[b]',
  morestring=[b]",
  alsoletter={:},
  alsodigit={-}
}

\lstdefinelanguage{JavaScript}{
  morekeywords={typeof, new, true, false, catch, function, return, null, catch, switch, var, if, in, while, do, else, case, break},
  morecomment=[s]{/*}{*/},
  morecomment=[l]//,
  morestring=[b]",
  morestring=[b]'
}

\lstdefinelanguage{HTML5}{
  language=html,
  sensitive=true,	
  alsoletter={<>=-},	
  morecomment=[s]{<!-}{-->},
  tag=[s],
  otherkeywords={
  >,
	<!DOCTYPE,
  </html, <html, <head, <title, </title, <style, </style, <link, </head, <meta, />,
	</body, <body,
	</div, <div, </div>, 
	</p, <p, </p>,
	</script, <script,
  <canvas, /canvas>, <svg, <rect, <animateTransform, </rect>, </svg>, <video, <source, <iframe, </iframe>, </video>, <image, </image>, <header, </header, <article, </article
  },
  ndkeywords={
  =,
  charset=, src=, id=, width=, height=, style=, type=, rel=, href=,
  fill=, attributeName=, begin=, dur=, from=, to=, poster=, controls=, x=, y=, repeatCount=, xlink:href=,
  margin:, padding:, background-image:, border:, top:, left:, position:, width:, height:, margin-top:, margin-bottom:, font-size:, line-height:,
  transform:, -moz-transform:, -webkit-transform:,
  animation:, -webkit-animation:,
  transition:,  transition-duration:, transition-property:, transition-timing-function:,
  }
}

\lstdefinestyle{htmlcssjs} {%
  backgroundcolor=\color{editorGray},
  basicstyle={\footnotesize\ttfamily},   
  frame=lines,
  xleftmargin={0.5cm},
  numbers=left,
  stepnumber=1,
  firstnumber=1,
  numberfirstline=true,	
  identifierstyle=\color{black},
  keywordstyle=\color{blue}\bfseries,
  ndkeywordstyle=\color{editorGreen}\bfseries,
  stringstyle=\color{editorOcher}\ttfamily,
  commentstyle=\color{brown}\ttfamily,
  language=HTML5,
  alsolanguage=JavaScript,
  alsodigit={.:;},	
  tabsize=2,
  showtabs=false,
  showspaces=false,
  showstringspaces=false,
  extendedchars=true,
  breaklines=true,
  literate=%
  {Ö}{{\"O}}1
  {Ä}{{\"A}}1
  {Ü}{{\"U}}1
  {ß}{{\ss}}1
  {ü}{{\"u}}1
  {ä}{{\"a}}1
  {ö}{{\"o}}1
}

\lstdefinelanguage{json}{
    numbers=left,
    numberstyle=\small,
    frame=single,
    rulecolor=\color{black},
    showspaces=false,
    showtabs=false,
    breaklines=true,
    postbreak=\raisebox{0ex}[0ex][0ex]{\ensuremath{\color{gray}\hookrightarrow\space}},
    breakatwhitespace=true,
    basicstyle=\ttfamily\small,
    upquote=true,
    morestring=[b]",
    stringstyle=\color{string},
    literate=
     *{0}{{{\color{numb}0}}}{1}
      {1}{{{\color{numb}1}}}{1}
      {2}{{{\color{numb}2}}}{1}
      {3}{{{\color{numb}3}}}{1}
      {4}{{{\color{numb}4}}}{1}
      {5}{{{\color{numb}5}}}{1}
      {6}{{{\color{numb}6}}}{1}
      {7}{{{\color{numb}7}}}{1}
      {8}{{{\color{numb}8}}}{1}
      {9}{{{\color{numb}9}}}{1}
      {\{}{{{\color{delim}{\{}}}}{1}
      {\}}{{{\color{delim}{\}}}}}{1}
      {[}{{{\color{delim}{[}}}}{1}
      {]}{{{\color{delim}{]}}}}{1},
}

\lstdefinestyle{C++} {%
  backgroundcolor=\color{editorGray},
  basicstyle={\footnotesize\ttfamily},   
  frame=lines,
  xleftmargin={0.5cm},
  numbers=left,
  stepnumber=1,
  firstnumber=1,
  numberfirstline=true,	
  identifierstyle=\color{black},
  keywordstyle=\color{blue}\bfseries,
  ndkeywordstyle=\color{editorGreen}\bfseries,
  stringstyle=\color{editorOcher}\ttfamily,
  commentstyle=\color{green}\ttfamily,
  morecomment=[l][\color{green}]{\%},
  language=C++,
  alsodigit={.:;},	
  tabsize=2,
  showtabs=false,
  showspaces=false,
  showstringspaces=false,
  extendedchars=true,
  breaklines=true,
  literate=%
  {Ö}{{\"O}}1
  {Ä}{{\"A}}1
  {Ü}{{\"U}}1
  {ß}{{\ss}}1
  {ü}{{\"u}}1
  {ä}{{\"a}}1
  {ö}{{\"o}}1
}

\lstdefinelanguage{PDF}{
  keywords={obj, endobj, trailer},
  morekeywords={Pages, Kids, Type, Action, Page, Parent, Count, OpenAction, S, JavaScript, JS, Size, Root},
  sensitive=false,
  comment=[l]\%,
  morecomment=[l]\%,
  morestring=[b]"
}

\lstdefinestyle{PDF}{
  language=PDF,
  backgroundcolor=\color{editorGray},
  basicstyle={\footnotesize\ttfamily},
  commentstyle=\color{olive}\ttfamily,
  keywordstyle=\color{blue}\bfseries,
  stringstyle=\color{orange}\ttfamily,
  emphstyle=\color{cyan},
  frame=lines,
  numbers=left,
  breaklines=true
}

\lstdefinelanguage{prompt}{
  keywords={},
  morekeywords={},
  sensitive=false,
}

\lstdefinestyle{prompt}{
  language=prompt,
  backgroundcolor=\color{editorGray},
  basicstyle={\footnotesize\ttfamily},
  frame=lines,
  numbers=left,
  breaklines=true
}

\newcommand{\ourtool}{{\textsc{PDFuzzer}}\xspace}

\newcommand{\func}[1]{\textit{#1}}

\begin{document}

\title{From Documentation to Zero-day Vulnerabilities:\\LLM-Driven Fuzzing of JavaScript Engines in PDF Readers}

\author{Suyue Guo}
\affiliation{%
  \institution{University of California, Santa Barbara}
  \city{Santa Barbara}
  \state{California}
  \country{USA}}
\email{sguo568@ucsb.edu}

\author{Stijn Pletinckx}
\affiliation{%
  \institution{University of California, Santa Barbara}
  \city{Santa Barbara}
  \state{California}
  \country{USA}}
\email{stijn@ucsb.edu}

\author{Tianle Yu}
\affiliation{%
  \institution{University of California, Santa Barbara}
  \city{Santa Barbara}
  \state{California}
  \country{USA}}
\email{tianleyu@stanford.edu}

\author{Yigitcan Kaya}
\affiliation{%
  \institution{University of California, Santa Barbara}
  \city{Santa Barbara}
  \state{California}
  \country{USA}}
\email{yigitcan@ucsb.edu}

\author{Saad Ullah}
\affiliation{%
  \institution{Boston University}
  \city{Boston}
  \state{Massachusetts}
  \country{USA}}
\email{saadu@bu.edu}

\author{Wenbo Guo}
\affiliation{%
  \institution{University of California, Santa Barbara}
  \city{Santa Barbara}
  \state{California}
  \country{USA}}
\email{henrygwb@ucsb.edu}

\author{Christopher Kruegel}
\affiliation{%
  \institution{University of California, Santa Barbara}
  \city{Santa Barbara}
  \state{California}
  \country{USA}}
\email{chris@cs.ucsb.edu}

\author{Giovanni Vigna}
\affiliation{%
  \institution{University of California, Santa Barbara}
  \city{Santa Barbara}
  \state{California}
  \country{USA}}
\email{vigna@ucsb.edu}

\renewcommand{\shortauthors}{Suyue Guo et al.}

\begin{abstract}

Existing fuzzers for PDF readers rely on simple test cases that involve only \textit{individual} API calls, leading to limited coverage and potentially missing vulnerabilities that require \textit{sequences} of API calls.
To address these limitations, we propose \ourtool, a novel PDF engine fuzzer that automatically generates complex and meaningful API call sequences.
\ourtool first uses a Large Language Model (LLM) to construct context-free grammars and infer the relationships between individual API calls from specifications extracted from JavaScript API manuals and execution traces.
Based on the grammars and relationships, \ourtool employs a constraint solver to generate concrete API call sequences for fuzzing.
Our experiments show that \ourtool significantly outperforms state-of-the-art PDF fuzzers (TypeOracle, Favocado, and Cooper) and LLM-based fuzzers (Fuzz4All, naive LLM) on three mainstream PDF readers: Adobe Acrobat Reader, Foxit PDF Reader, and PDF-XChange Editor.
\ourtool achieves up to 48\% higher coverage than existing tools and identifies 31 zero-day vulnerabilities in these readers, from information leakage to arbitrary code execution.
Our ablation study validates the necessity of each component, including LLMs, which achieve high accuracy across all pipeline stages (93-98\%). 
We disclosed all vulnerabilities to the vendors via a coordinated vulnerability disclosure process and received bug bounties.
\end{abstract}



\maketitle

\section{Introduction}

The Portable Document Format (PDF)'s support for JavaScript code offers interactive capabilities, where the code is parsed and executed by a dedicated JavaScript engine embedded in the PDF reader.
Because this engine runs user-provided code on a host system, it is crucial to identify the vulnerabilities that could lead to malicious code execution.
To achieve this, existing efforts~\cite{DBLP:conf/icse/typeoracle, DBLP:conf/ndss/cooper, DBLP:conf/ndss/favocado} generate JavaScript-based test cases to fuzz PDF readers. 
However, these approaches suffer from significant limitations in both automation and relationship inference. 
Specifically, Favocado~\cite{DBLP:conf/ndss/favocado} and Cooper~\cite{DBLP:conf/ndss/cooper} require substantial manual effort to analyze natural-language API documentation and construct machine-readable test-generation rules. 
TypeOracle~\cite{DBLP:conf/icse/typeoracle}, although automated, relies on simple name-based similarity for relationship inference, missing complex semantic dependencies between API calls.
The challenge is compounded by the fact that over 30\% of PDF reader functions are undocumented~\cite{DBLP:conf/icse/typeoracle}, creating substantial gaps in API coverage. 
Even tools that attempt to model API relationships~\cite{DBLP:conf/ndss/favocado} only capture basic producer-consumer dependencies, in which one API function's return value directly serves as another's input parameter. 

Specifically, by systematically examining Adobe's JavaScript API documentation~\cite{adobe_acrobat_manual, adobe_acrobat_manual2}, we identify two relationship types that producer-consumer-only approaches cannot represent.
The first is value-constraint relationships, where parameters of two API calls must satisfy equality/inequality constraints (the parameter names need not match; e.g., \func{addField} and \func{removeField} require the same \func{cName} with a specific ordering, while \func{popUpMenuEx}'s \func{cName} and \func{search.query}'s \func{cQuery} must hold the same string).
The second is implicit relationships, in which APIs interact through shared state rather than explicit parameters (e.g., \func{setAction} registers handlers that are later triggered by \func{setFocus}).
The combination of incomplete API coverage and limited relationship inference creates a substantial gap in testing effectiveness, as both undocumented APIs and complex interaction patterns that could reveal vulnerabilities remain largely unexplored.

In this work, we propose \ourtool, which leverages LLMs and constraint solvers to automatically infer comprehensive specifications for both documented and undocumented APIs and to capture intricate relationships among API calls, thereby generating high-quality (sequential) inputs. 
\ourtool systematically identifies and models three types of API relationships---producer-consumer, value-constraint, and implicit---while automatically inferring detailed specifications for the substantial portion of undocumented functionality, enabling comprehensive testing that existing approaches cannot achieve.
The key distinction is that \ourtool does not stop at producer-consumer relationships.
It turns natural-language API semantics into symbolic constraints that capture value coupling, ordering requirements, and state-dependent interactions between API calls.
Technically, we first extract comprehensive API specifications through a dual-path approach that handles both documented and undocumented APIs. For documented APIs, we directly extract structured information from official manuals using regular expressions.
For undocumented APIs, we extract their signatures via differential analysis of execution traces~\cite{DBLP:conf/icse/typeoracle} and leverage LLMs to infer comprehensive specifications.
Here, LLMs capture API design patterns, naming conventions, and parameter semantics to transform minimal signature information into detailed specifications that match the quality of documented APIs, outperforming signature-based approaches.
Using these specifications, we create context-free grammars (CFGs) for each API function and its parameters, providing a structured syntax for generating syntactically valid API function invocations.
We then use LLMs to infer inter-API relationships, leveraging their ability to perform multi-step reasoning over diverse documentation and previously inferred specifications~ \cite{10.5555/3600270.3602070}. 
This allows us to move beyond simple type matches to identify the dependencies that govern realistic API call sequences.
To ensure precision, we express the relationship extracted by LLMs as constraints for pairs of API calls, and use SMT solvers (Z3) to solve the constraints to generate concrete test cases with \emph{sequences} of API calls.

We compared \ourtool with three state-of-the-art PDF reader fuzzers (TypeOracle~\cite{DBLP:conf/icse/typeoracle},  Favocado~\cite{DBLP:conf/ndss/favocado}, and Cooper~\cite{DBLP:conf/ndss/cooper}) and two LLM-based methods for general-purpose fuzzing (Fuzz4All~\cite{DBLP:conf/icse/fuzz4all} and a naive LLM method based on TitanFuzz~\cite{DBLP:conf/issta/titanfuzz}).
We evaluated these methods on three popular PDF readers, Adobe Acrobat Reader, Foxit PDF Reader, and PDF-XChange Editor, using two experiments:
(1) tracking coverage over 24 hours and (2) recording vulnerabilities found over two weeks.
\ourtool reaches up to 48\% higher coverage and discovers 31 zero-day vulnerabilities across three targets, compared to at most 6 found by competing tools.
These vulnerabilities include null pointer dereferences, buffer overflows, and use-after-free bugs, with potential for arbitrary code execution. 
%
We coordinated disclosure of all vulnerabilities with the vendors; 26 have been confirmed or fixed, yielding \$2,450 in bug bounties.


We validated our design choices through a comprehensive ablation study on each key component of \ourtool, by either turning it off or replacing it with the appropriate baseline technique, such as TypeOracle~\cite{DBLP:conf/icse/typeoracle}.
Our evaluation demonstrates that each component contributes significantly to the overall effectiveness:
undocumented API specification inference achieves up to 28\% higher coverage than type-only approaches, 
parameter-level grammar generation provides up to 18\% higher coverage than API function-level grammar generation, 
strong symbolic relationship modeling improves coverage by up to 8\% compared to candidate (co-occurrence-only) relationships, and value-constraint and implicit relationships specifically add up to 15.5\% over a producer-consumer-only configuration,
and PDF objects integration enhances coverage by up to 8\% beyond JavaScript-only approaches.
Additionally, spot-checks of the LLM-driven components across our pipeline showed consistent reliability, with 93–98\% accuracy in API specification extraction, grammar generation, and relationship inference.

\smallskip
\noindent In summary, our main contributions are as follows: \\[-0.3cm]

\noindent $\bullet$ We propose \ourtool, an LLM-based fuzzer designed to achieve deeper program logic coverage in PDF readers. 
It generates test inputs by extracting detailed specifications for undocumented API calls and leveraging LLMs to infer symbolic relationships between JavaScript functions.
Unlike producer-consumer-only approaches, \ourtool also infers value-constraint and implicit relationships from natural-language API specifications, enforcing them through SMT solving during test generation.

\noindent $\bullet$  We experimentally demonstrate that \ourtool achieves up to 48\% higher code coverage and discovers 31 zero-day vulnerabilities in three popular PDF readers (Adobe Acrobat Reader, Foxit PDF Reader, and PDF-XChange Editor), significantly outperforming state-of-the-art traditional and LLM-based fuzzers.

\noindent $\bullet$ We validate our design through detailed ablation studies on key components, including specification inference, CFG generation, and relationship inference, showing how each component contributes to the overall effectiveness.

\section{Background and Existing Work}
\label{sec:bg}

\subsection{JavaScript Engines in PDF Readers}
PDF documents can embed JavaScript code to enable interactive features such as spell checking and form submission.
To support this functionality, PDF readers include scripting engines that parse and execute the embedded code.
Beyond standard JavaScript, these engines expose specialized APIs that invoke native code within the reader, altering how documents are rendered.
The APIs supported by PDF readers are documented in vendor API manuals~\cite{adobe_acrobat_manual, adobe_acrobat_manual2}.
While powerful, this functionality also introduces risks: adversaries have exploited vulnerable API functions, such as Adobe's \func{util.printf} (buffer overflow~\cite {CVE-2008-2992}) and Foxit's \func{app.launchURL} (remote code execution~\cite{CVE-2017-10951}) to compromise systems.
Thorough testing of PDF readers is therefore critical.
In particular, dynamic testing requires generating valid PDF documents with realistic JavaScript API calls, a task made challenging by the need to interpret API manuals and produce valid, mutated API calls that are embedded into executable PDF files.
Throughout this paper, we use \emph{API specification} to denote the information needed to generate and validate such an API call: its owning object (the JavaScript object that exposes the API, e.g., \func{app} in \func{app.popUpMenuEx}), function or property name, parameters, constraints, return value, and behavior.

\subsection{Threat Model}

We consider a malicious attacker who crafts a PDF document with embedded JavaScript in order to trigger vulnerabilities in the PDF reader's JavaScript engine.
The attack is realized when a victim opens the malicious PDF, causing the embedded script to execute inside the reader and potentially leading to memory corruption, information leakage, denial of service, or arbitrary code execution on the victim's system.
This threat is realistic and consistent with prior in-the-wild cases of JavaScript-based PDF exploitation, such as CVE-2024-28888~\cite{CVE-2024-28888} and CVE-2024-34099~\cite{CVE-2024-34099}.
Accordingly, our goal is to generate high-quality JavaScript-based PDF test cases that can expose such vulnerability-triggering behaviors in mainstream PDF readers.

\subsection{Existing PDF Fuzzers and Limitations}
\label{sec:background:existing_work}

Due to the closed-source nature of mainstream PDF readers, static analysis is not commonly used for the security evaluation of JavaScript engines.
As such, existing work on analyzing the security of JavaScript engines in PDF readers has focused mostly on dynamic testing techniques, more specifically, fuzzing~\cite{DBLP:conf/icse/typeoracle, DBLP:conf/ndss/favocado, DBLP:conf/ndss/cooper}. 
To perform fuzzing on PDF readers, one needs to embed in PDF documents JavaScript code that invokes a large and diverse set of JavaScript API calls (this is called the fuzzing corpus).
One can then input these PDF documents into the PDF reader to observe their behavior.
This basic approach is used by all prior work~\cite{DBLP:conf/icse/typeoracle, DBLP:conf/ndss/favocado, DBLP:conf/ndss/cooper}, but it suffers from three key limitations:


\noindent \ding{192} {\bf Missing undocumented API specifications.}
Favocado~\cite{DBLP:conf/ndss/favocado} and Cooper~\cite{DBLP:conf/ndss/cooper} miss a wide range of potential test cases because they rely solely on official API manuals.
TypeOracle~\cite{DBLP:conf/icse/typeoracle} can identify undocumented API functions via differential analysis but fails to extract details about their functionality, preventing the generation of high-quality (valid) invocations for these undocumented API functions.

\noindent \ding{193} {\bf Failure to capture complex inter-API relationships.}
TypeOracle~\cite{DBLP:conf/icse/typeoracle} uses differential analysis on execution traces to infer parameter type information for API functions and finds inter-API relationships through name similarity.
Other API fuzzing approaches, including Favocado~\cite{DBLP:conf/ndss/favocado}, focus on producer-consumer relationships, where the return value of one API is used as an input parameter to another API.
Both name-based matching and producer-consumer matching are useful, but they are insufficient for relationships whose validity depends on concrete parameter values or shared program state.
For example, while \func{addField} and \func{removeField} both have a \func{cName} parameter type, they are related (connected) only if the \func{cName} value is the same.
Additionally, the order of these API calls is also important - \func{addField} needs to be called before \func{removeField} (and with the same parameter) for them to have a relationship.
This illustrates a value-constraint relationship, while implicit relationships arise from shared reader state; Section~\ref{sec:relation_definition} defines both cases in detail.


\noindent \ding{194} {\bf Lack of automated specification extraction.}
While existing PDF reader fuzzers automate the test case generation and execution process, they still require substantial manual effort to convert natural language API documentation into machine-readable test generation rules.
For instance, Favocado~\cite{DBLP:conf/ndss/favocado} demonstrated that even with automated parsing pipelines, significant human expertise is still necessary to manually review hundreds of pages of API documentation, interpret semantic relationships between API functions, and construct comprehensive specification rules for JavaScript APIs.
In practice, human experts still need to fill gaps left by parsers.
They supplement missing fields, correct extraction errors, and add initialization code for binding objects (internal PDF objects that bridge JavaScript APIs to native reader functions) when environment-specific data is required~\cite{DBLP:conf/ndss/favocado}.
Cooper~\cite{DBLP:conf/ndss/cooper} similarly depends on manual analysis of API documentation to create test generation templates.
TypeOracle~\cite{DBLP:conf/icse/typeoracle} automates API signature extraction via differential analysis, but cannot infer semantic descriptions and behavioral constraints from documentation, limiting its ability to generate high-quality test cases.
With Adobe's official API manual~\cite{adobe_acrobat_manual, adobe_acrobat_manual2} exceeding 700 pages of natural language descriptions, manual specification extraction remains time-consuming, labor-intensive, and often incomplete.

\section{Preliminary: Inter-API Relationships}
\label{sec:relation_definition}

Effective fuzzing of PDF readers requires understanding interactions between JavaScript API functions.
Existing fuzzers largely generate isolated API calls; however, many vulnerabilities are only triggered through orchestrated sequences of related calls.
We use the term \emph{inter-API relationship} to refer to a semantic dependency between two API calls that affects whether they should appear together, in which order they should execute, or what constraints their parameter values or shared state must satisfy.
Inter-API relationships come in two strength tiers.
Candidate relationships are the weaker tier: a coarse co-occurrence hypothesis encoding only that two API functions may meaningfully appear together, without specific parameter or ordering constraints.
Strong symbolic relationships are the stronger tier: confirmed relationships that additionally carry symbolic constraints over parameter values or shared state, and optionally an ordering requirement (Section~\ref{sec:method} details how each tier is inferred).
From Adobe's JavaScript API documentation~\cite{adobe_acrobat_manual, adobe_acrobat_manual2} and prior vulnerability-triggering test cases~\cite{DBLP:conf/icse/typeoracle, DBLP:conf/ndss/favocado, DBLP:conf/ndss/cooper}, we identify three categories of strong symbolic relationships.
Producer-consumer relationships capture the type-based data flow commonly modeled by prior API fuzzing work~\cite{DBLP:conf/ndss/favocado, DBLP:conf/icse/restler}; the other two categories, value-constraint and implicit relationships, are the main types that let \ourtool go beyond producer-consumer-only approaches.

\subsection{Producer-Consumer Relationship}
A producer-consumer relationship connects $\mathrm{API}_1$'s return value to an argument of $\mathrm{API}_2$, so the generated sequence must execute $\mathrm{API}_1$ before $\mathrm{API}_2$ and pass forward $\mathrm{API}_1$'s return value.

\begin{lstlisting}[caption={Producer-Consumer Relationship Example}\label{lst:producer_consumer_case},style=htmlcssjs,showspaces=false, showstringspaces=false]
var myIcon = Doc.getIcon({cName: "iconName"});
Field.buttonSetIcon({oIcon: myIcon, nFace: 0});
\end{lstlisting}

\noindent \func{Doc.getIcon} returns an Icon object representing a named icon in the document, which \func{Field.buttonSetIcon} then consumes as its \func{oIcon} parameter to set a button's icon.
The dependency is characterized by type compatibility between the producer's return value and the consumer's input parameter.

\subsection{Value-Constraint Relationship}
A value-constraint relationship is a constraint over parameters from two API calls -- such as equality, ranges, or set-membership -- that must hold for the calls to interact meaningfully.

\begin{lstlisting}[caption={Value-Constraint Relationship Example}\label{lst:value_constraint_case},style=htmlcssjs,showspaces=false, showstringspaces=false]
Doc.addField({cName: "myField", cFieldType: "text", ...});
Doc.getField({cName: "myField"});
\end{lstlisting}

\noindent Both \func{Doc.addField} and \func{Doc.getField} take a \func{cName} parameter specifying the field name.
For meaningful (correct) behavior, the \func{cName} values must be the same (that is, the function arguments refer to the same field).
Also, \func{addField} must precede \func{getField}, since a field must exist before it can be retrieved.
Unlike producer-consumer, the dependency here is on the actual {\it values} of two input parameters rather type compatibility between an input and a return value.
Moreover, the constraints can be broader that equality and capture more complex relationships.

\subsection{Implicit Relationship}
An implicit relationship arises when two API functions interact via shared system state or objects rather than explicit parameter connections, with no direct argument-to-argument or return-to-argument data flow.

\begin{lstlisting}[caption={Implicit Relationship Example}\label{lst:implicit_case},style=htmlcssjs,showspaces=false, showstringspaces=false]
Field.setAction({cTrigger: "OnFocus", cScript: "some_action"});
Field.setFocus();
\end{lstlisting}

\noindent \func{Field.setAction} configures the action to trigger when a specific event (named by its \func{cTrigger} parameter) occurs on a field, while \func{Field.setFocus} transfers keyboard focus to the field and invokes the corresponding event handler.
When \func{setAction} is called with \func{cTrigger}~$=$~\texttt{"OnFocus"} before \func{setFocus}, the focus event dispatches the previously configured action.
The dependency runs through the field's shared internal state -- \func{setAction} writes the action configuration and \func{setFocus} reads it -- rather than through any direct argument-to-argument or return-to-argument data flow, and is activated only when \func{cTrigger} matches the event the second call fires.

\section{Methodology}
\label{sec:method}

We propose \ourtool, an LLM-based fuzzing tool for JavaScript engines in PDF readers that generates higher-quality, complex, and semantically meaningful test cases than prior work, achieving greater code coverage and vulnerability discovery. \ourtool systematically addresses the three key limitations of existing PDF reader fuzzers through a principled design that leverages LLMs and constraint solvers.

\subsection{Overview}
\label{sec:overview}

\begin{figure*}[t!]
    \centering
    \makebox[\linewidth][c]{\resizebox{1.06\linewidth}{!}{\input{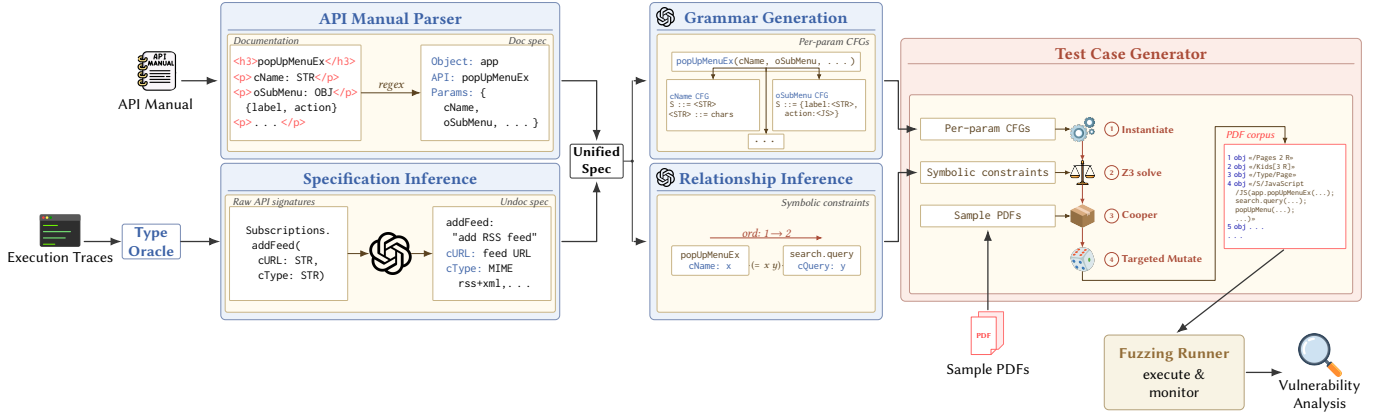}}}
    \caption{Overview of \ourtool.
    Documented APIs are extracted by the API Manual Parser; undocumented APIs are recovered by Specification Inference from execution traces.
    Both \emph{Doc spec} and \emph{Undoc spec} feed Grammar Generation and Relationship Inference in parallel.
    Their outputs (per-parameter CFGs and symbolic constraints) drive the Test Case Generator, which additionally consumes Sample PDFs for native-object templates.
    The resulting PDF corpus is exercised by the Fuzzing Runner and triaged by Vulnerability Analysis.}
    \label{fig:workflow}
\end{figure*}

\ourtool comprises four main components that collaboratively generate high-quality API call sequences (see Figure~\ref{fig:workflow}).

\noindent \textbf{API Specification Extraction.}
The first component extracts API specifications from two sources: documented APIs via our API Manual Parser over the official manuals~\cite{adobe_acrobat_manual, adobe_acrobat_manual2}, and undocumented APIs via differential analysis of execution traces (which yields only minimal signatures \func{<object>.<method>(<param>: <type>, ...)}). The Specification Inference module then uses an LLM to upgrade these signatures into comprehensive specifications with descriptions and parameter constraints, addressing Limitation~1.

\noindent \textbf{Grammar Generation.}
This component compiles each specification into context-free grammars (CFGs) for individual API functions and their parameters via a two-phase, parameter-level approach: each parameter's grammar is generated independently, preserving fine-grained constraints (enumerated value sets, format requirements) that a function-level grammar would overgeneralize. This addresses part of Limitation~3.

\noindent \textbf{Relationship Inference.}
Operating in parallel with Grammar Generation, this component identifies inter-API dependencies (Limitation~2) via a two-stage LLM process with Retrieval Augmented Generation (RAG)~\cite{rag}: Stage~1 selects candidate API pairs that may co-occur meaningfully; Stage~2 promotes a subset to strong symbolic relationships, encoding parameter constraints as SMT-LIB2~\cite{smt} expressions and recording execution ordering. This separates semantic reasoning (LLM) from constraint solving (SMT).

\noindent \textbf{Test Case Generator.}
The final component synthesizes concrete PDF inputs by instantiating each API from its CFG, sequencing related calls under candidate relationships, and using a Z3~\cite{z3-solver} SMT solver to satisfy strong symbolic constraints. We additionally integrate Cooper's~\cite{DBLP:conf/ndss/cooper} cooperative mutation to populate the necessary PDF native structures and apply targeted mutations to a subset of test cases for robustness probing, completing Limitation~3.

\subsection{Running Example}
\label{sec:running_example}
We use a single end-to-end example, traced through the rest of this section's component subsections and revisited in our vulnerability case study (Section~\ref{sec:eval_vulnerability}), to illustrate how the four components above turn API documentation and execution traces into vulnerability-triggering inputs.
The example, shown in Listing~\ref{lst:case2}, is a test case \ourtool generated and used to trigger a use-after-free vulnerability in Foxit PDF Reader (vulnerability ID~5 in Table~\ref{tab:vulnerabilities}, later fixed and assigned a CVE entry).

\begin{lstlisting}[caption={Running example: \ourtool-generated test case triggering a use-after-free vulnerability in Foxit PDF Reader.}\label{lst:case2},style=htmlcssjs,showspaces=false, showstringspaces=false]
app.popUpMenuEx({cName: "popup", oSubMenu: {label: "kM", action: "mFv"}});
search.query({cQuery: "popup"});
app.popUpMenu({cItem: "M", Array: ["21"]});
\end{lstlisting}

This test case simultaneously exhibits two structural features that prior fuzzers cannot capture together.
First, Lines~1-2 form a \emph{strong symbolic relationship of type value-constraint}: the \func{cName} of \func{app.popUpMenuEx} and the \func{cQuery} of \func{search.query} must hold the same string for the memory allocation in \func{app.popUpMenuEx} and the deallocation in \func{search.query} to operate on the same memory region.
Second, Line~3 (\func{app.popUpMenu}) is associated with Line~1 only via \emph{candidate co-occurrence}: it contributes a related API call to the sequence but has no parameter-level coupling, so its arguments are generated independently.
\ourtool models these two tiers explicitly: candidate relationships specify which API functions can co-occur in a sequence, and strong symbolic relationships further enforce parameter-level constraints (encoded as SMT-LIB2~\cite{smt} expressions, the standard input format consumed by SMT solvers such as Z3~\cite{z3-solver}) and execution ordering.
The remaining subsections follow this example through specification extraction, parameter-level grammar generation, two-stage relationship inference, and final test-case generation, with a closing ``Running Example'' paragraph in each subsection showing how its output shapes the listing (test case) above.

TypeOracle's~\cite{DBLP:conf/icse/typeoracle} name-similarity matching would link \\ \func{app.popUpMenuEx} only to \func{app.popUpMenu} (which share the \texttt{popUpMenu} prefix) and never to \func{search.query} (whose name has no overlap), so it cannot enforce the required \func{cName}~$=$~\func{cQuery}.
Favocado~\cite{DBLP:conf/ndss/favocado} and RESTler-style~\cite{DBLP:conf/icse/restler} producer-consumer matching cannot connect \func{app.popUpMenuEx} and \func{search.query} either, since neither call returns a value consumed by the other.
The bug therefore remains out of reach for prior fuzzers.

\subsection{API Manual Parser and Specification Inference}
The JavaScript API manual is an HTML document with a structured enumeration of the available function calls (and their parameters) for the JavaScript engine in PDF readers. 
The manual lists Adobe's built-in objects and the API functions that objects can call, including their properties. 
We evaluate three PDF readers: Adobe Acrobat Reader, Foxit PDF Reader, and PDF-XChange Editor.
Because Foxit and PDF-XChange do not provide comprehensive API manuals, we use Adobe's API manual as the reference specification: Adobe's JavaScript API has become the de facto standard for PDF JavaScript engines, and Foxit and PDF-XChange both implement it, as noted in prior work~\cite{DBLP:conf/icse/typeoracle}.
For each API function, the documentation first provides a paragraph explaining its functionality. 
The manual then lists the parameter names for the API function.
Notably, for most parameters, the documentation does not explicitly state what type is required.
As a result, simply parsing the documentation is not enough, and we must infer these parameter types for building API invocation rules.
If the API function has a return value, the documentation also includes a description of the return value.

We extract specifications for each API function by using regular expressions to match HTML objects and capture hierarchical information.
Specific titles identify each API function’s content area, which we parse into JSON files. 
Extracted data includes the object that owns the function, function name, type (method or property), description, parameters, parameter descriptions, and return value.


\smallskip \noindent \textbf{Specification Inference.}
PDF readers contain numerous undocumented API functions that are not described in official manuals. 
These may be internal-only functions not intended for external use, legacy compatibility functions, or vendor-specific extensions. 
Fuzzing such calls is known to be an effective method for vulnerability discovery, as they often receive less security review and may lack proper input validation~\cite{afgen}.

TypeOracle~\cite{DBLP:conf/icse/typeoracle} can infer parameter types via differential analysis of execution traces, but cannot capture semantic functionality or behavioral constraints. We close this gap by combining TypeOracle's type inference with LLM-based semantic specification generation in three stages.

\noindent \textit{Stage~1: Type Extraction.}
We apply TypeOracle's differential analysis to obtain raw API signatures of the form \func{<object>.<method>(<param1>: <type1>, ...)}---e.g., \func{Doc.getField(cName: String)}---by extracting object/method names from internal call tables and inferring parameter types from how instruction operands vary across runs (e.g., a length-related operand changing from 2 to 4 when passing \texttt{"YY"} vs.\ \texttt{"zzzz"} indicates a String parameter). These signatures reveal only the structural skeleton; they omit each parameter's role, valid ranges, the function's purpose, return-value semantics, and side effects.

\noindent \textit{Stage~2: Context Preparation.}
For each undocumented API, we collect contextual information from the structured JSON specs produced by our \emph{API Manual Parser}.
This context includes (1) the parent object's description and functionality, (2) other APIs (documented and undocumented) on the same object, and (3) the parameter names and types inferred in Stage~1.

\noindent \textit{Stage~3: LLM-based Specification Generation.}
We prompt the LLM to generate comprehensive specifications using contextual information from Stage~2.
Importantly, this applies only to undocumented APIs; documented APIs are handled by our API Manual Parser with regular expressions, since their specifications already exist in the official manuals in natural language.
The LLM uses its understanding of API design patterns and naming conventions to infer: 
(1) functional descriptions for the API function and each parameter, 
(2) parameter value constraints and valid ranges, 
(3) return value specifications, and 
(4) behavioral semantics.
For example, given the raw signature \func{Subscriptions.addFeed(cURL: String, cType: String)}, the LLM reasons that the parent \func{Subscriptions} object handles feed subscriptions, that \func{cURL} carries a feed URL, and that \func{cType} must be a MIME/content-type string with concrete values such as \texttt{"application/rss+xml"} or \texttt{"application/json"}.
This inference is drawn from the API name, parameter names, and parent-object context alone, since the \func{Subscriptions} object has no entry in Adobe's manual and no documented counterpart can serve as a template.
TypeOracle alone yields only \texttt{\{cURL: String, cType: String\}}, insufficient for invocations a feed engine can parse.

These outputs are expressed in natural language, consistent with Adobe's API documentation, which primarily consists of prose descriptions rather than formal constraints.

The LLM generates specifications in a structured JSON format consistent with documented APIs, ensuring pipeline uniformity for both types.
The JSON output includes fields requiring the LLM to explain its reasoning, enabling us to interpret and validate its inferences during accuracy evaluation (Section~\ref{sec:llm_accuracy}).

\vspace{0.05cm} \noindent \textbf{Error Handling.}
To handle potential LLM errors, we implement JSON format verification.
Our error recovery strategy automatically retries failed attempts up to a specific number of times, incorporating specific error feedback into subsequent prompts (e.g., ``Previous JSON parsing failed: [error]. Please generate a valid JSON format'').

\smallskip \noindent \textbf{Running Example.}
All three APIs in Listing~\ref{lst:case2} (\func{app.popUpMenuEx}, \func{search.query}, and \func{app.popUpMenu}) are documented in Adobe's manual and therefore traverse the Manual Parser path, producing JSON specifications listing each API's parameters (e.g., \func{popUpMenuEx} owns \func{cName} and the nested \func{oSubMenu}).
The Specification Inference path is exercised by undocumented APIs.

\subsection{Grammar Generation}

After extracting API specifications for documented and undocumented API functions, we have structured JSON specifications containing natural-language descriptions of all the API functions that the JavaScript engine makes available. 
Each JSON specification follows a hierarchical format where API function parameters are organized with their individual descriptions and types.
However, systematic and automated test case generation requires these specifications in a machine-readable format suitable for programmatic processing. 
To achieve this, we leverage context-free grammars (CFGs), drawing inspiration from grammar-based fuzzing approaches~\cite{DBLP:conf/ndss/nautilus, DBLP:conf/sigsoft/grammarinator} that use CFGs to automatically generate syntactically valid test inputs.
CFGs provide an ideal representation for API call statements and parameter generation rules due to their structured, parseable syntax and ability to capture complex constraints. 
Our initial approach involved directly feeding complete API function specifications to an LLM to generate comprehensive CFGs. 
While this method works effectively for simple API functions with few parameters (such as property getters and setters), it struggles with complex API functions with numerous parameters (as discussed below).

\smallskip \noindent \textbf{Challenge: Information Overload in Complex APIs.} 
The challenge became apparent with API functions like \func{Doc.submitForm}, which requires 23 distinct parameters. 
Consider the \func{cSubmitAs} parameter: although recognized as \func{String} type, it accepts only six specific values ({\tt FDF}, {\tt XML}, {\tt XDF}, etc.) corresponding to supported file formats. 
Processing such complex specifications in a single prompt causes information overload for LLMs. 
Consequently, this leads to overgeneralized grammars that generate random strings, failing to respect the constrained value sets.
This loss of granularity occurs because LLMs struggle to prioritize critical parameter-specific constraints when processing lengthy specifications within fixed context windows.

\smallskip \noindent \textbf{Solution: Two-Phase Parameter-Level Grammar Generation.} 
To address this challenge and capture fine-grained parameter constraints, we design a two-phase approach:

\textit{Phase 1: Specification Decomposition.} 
We parse the structured JSON format to systematically isolate each parameter's natural language description from the complete API specification.
For an API with $n$ parameters, this decomposition creates $n$ independent specification units, each preserving parameter-specific constraints, value restrictions, and type requirements. 
This decomposition ensures that each parameter's unique constraints receive focused attention during grammar generation.

\textit{Phase 2: Targeted Grammar Synthesis.} 
Each parameter specification unit is processed by the LLM using specialized prompts that include essential metadata, while maintaining focus on the specific parameter. 
The metadata includes API type (distinguishing between methods and properties with different invocation syntaxes), parent object, parameter name, and description.
The prompts incorporate detailed sample JSON grammar structures demonstrating proper BNF-style rule decomposition and enforce strict syntactic rules, including character escaping, complete symbol definition, and adherence to value generation constraints (e.g., enumerated string sets). 

This parameter-focused approach captures syntactic patterns and semantic constraints that would otherwise be missed in comprehensive function-level processing. 
The resulting grammars encode parameter-specific requirements, such as constrained value sets, numeric ranges, and format specifications, leading to higher-quality test case generation that respects API semantics while maintaining syntactic validity.

\smallskip \noindent \textbf{Error Handling and Normalization.}
Despite careful prompt engineering, CFG-generated API parameters sometimes violate JavaScript engine syntax requirements. 
We observed four recurring formatting issues that cause parsing failures in target readers.
These include missing quotes for string literals, leading zeros in numeric values (e.g., 007), unescaped quotes inside strings, and embedded newline characters.

To address these inconsistencies, we implement a lightweight Python post-processing step that applies type-specific correction rules during test case generation:

\textit{String normalization:} Removes any existing outer quotation marks, strips all internal quotation marks and newline characters from the string content, then wraps the cleaned content with proper quotation marks.

\textit{Number normalization:} Removes leading zeros from the integer portion while preserving the sign and any fractional components (e.g., \texttt{007.5} becomes \texttt{7.5}, \texttt{-0012} becomes \texttt{-12}).

\textit{Array normalization:} Removes outer square brackets, parses individual elements, applies normalization recursively to each element based on its inferred type, then reassembles the array with proper bracket notation.

\textit{Object normalization:} Strips outer curly braces, parses key-value pairs, ensures object keys begin with an alphabetic character (prepending a random letter if needed), recursively normalizes values, and reconstructs the object with proper braces.

This normalization layer ensures that all generated parameters conform to JavaScript syntax requirements before being embedded into PDF test cases, preventing test case rejection due to basic syntactic violations while preserving the semantic intent of the original parameter values.

Finally, to ensure cross-compatibility, we avoid generating extended Unicode syntax (\texttt{\textbackslash u\{XXXX\}}) in test cases, which is supported by Foxit and PDF-XChange but not by Adobe Acrobat Reader.

\smallskip \noindent \textbf{Running Example.}
For \func{app.popUpMenuEx} in Listing~\ref{lst:case2}, parameter-level decomposition produces independent CFGs for \func{cName} (a character-level string grammar) and \func{oSubMenu} (a nested-object grammar with separate rules for the \func{label} and \func{action} fields).
A function-level grammar would tend to flatten \func{oSubMenu} into an arbitrary value and lose the nested-field structure required by the target PDF reader's JavaScript engine.

\subsection{Relationship Inference}
The previous step of \ourtool results in grammars generated for individual API functions and their parameters.
However, many software vulnerabilities can only be triggered when multiple related API functions are called in a specific sequence~\cite{DBLP:conf/ndss/favocado, DBLP:conf/sosp/healer, DBLP:conf/uss/Fleischer0BBLPK23}.
Therefore, we need a method to infer relationships between API functions so that we can generate sequences of related API calls during test case generation.

We group inter-API relationships into two strength tiers: candidate relationships and strong symbolic relationships.
A candidate relationship is a coarse co-occurrence hypothesis: two API functions may be meaningful in the same sequence, but no order, argument mapping, or parameter constraint has been established.
A strong symbolic relationship is a confirmed relationship that records related arguments or return values, symbolic variables, an SMT-LIB2 constraint, an optional ordering requirement, and one of the semantic relationship types defined in Section~\ref{sec:relation_definition}.

Analyzing all $\binom{n}{2}=221{,}445$ pairwise relationships among $n=666$ APIs at $\sim$30~s of LLM analysis each would take over 1,845 hours and is intractable. We instead use a two-stage approach that turns this $O(n^2)$ problem into $O(n) + O(k^2)$ with $k \ll n$: Stage~1 uses fast RAG queries to surface candidate pairs (linear in $n$); Stage~2 runs expensive constraint analysis only on those candidates.

\smallskip \noindent \textbf{Stage~1: Candidate Relationship Extraction.}
We iterate over each API and prompt the LLM to identify related APIs from our specification database, stored in a Vector Store and queried via RAG~\cite{rag} to avoid context-window overflow (full prompt is available in our public artifact). The output is a set of candidate API pairs only; producer-consumer, value-constraint, and implicit category assignment happens in Stage~2.

\smallskip \noindent \textbf{Stage~2: Strong Relationship Analysis and Symbolization.}
As introduced in Section~\ref{sec:relation_definition}, each strong symbolic relationship is assigned one semantic type: Producer-Consumer, Value-Constraint, or Implicit.
For each candidate API pair, we use a single comprehensive prompt to analyze their specifications and determine if constraint-based relationships exist.
The prompt performs zero-shot analysis, instructing the LLM to simultaneously accomplish three key tasks in one inference pass:
(1) \textit{Parameter Symbolization}:  assign symbolic variables (e.g., x, y) to related parameters or return values,
(2) \textit{Constraint Generation}: express the relationship as SMT-LIB2 constraints (e.g., (= x y) for equality), and 
(3) \textit{Sequence Analysis}: determine if the API functions must be executed in a specific order.

The LLM returns a structured JSON object containing the API function names, parameter mappings, symbolic variables, execution sequence requirements, parameter types, and the SMT-LIB2 constraint expression.
For instance, when analyzing \func{Doc.addField} and \func{Doc.getField}, the LLM identifies that their \func{cName} parameters must be equal and that \func{addField} must precede \func{getField}.
The constraint \texttt{(= x y)} captures the parameter equality requirement, while the sequence flag indicates the execution dependency.


\smallskip \noindent \textbf{Integration with Test Generation.}
The inferred relationships are stored as constraint templates that guide the Test Case Generator.
During test sequence construction, candidate relationships determine which API functions can be combined, while strong symbolic relationships provide the constraints that an SMT solver must satisfy when generating concrete parameter values.
This two-tier approach enables efficient generation of complex, semantically meaningful API call sequences.

\smallskip \noindent \textbf{Running Example.}
For Listing~\ref{lst:case2}, Stage~1 returns two candidate pairs: \func{(app.popUpMenuEx, search.query)} and \func{(app.popUpMenuEx, app.popUpMenu)}.
Stage~2 promotes the first to a strong symbolic relationship of type value-constraint with parameter symbols \\ \func{app.popUpMenuEx.cName}~$\mapsto x$, \func{search.query.cQuery}~$\mapsto y$, constraint \texttt{(=\ x\ y)}, and ordering \func{app.popUpMenuEx} $\rightarrow$ \func{search.query}; the second pair remains a candidate, contributing a related API call without parameter-level coupling.

\subsection{Test Case Generator}
The Test Case Generator synthesizes concrete PDF test inputs by integrating outputs from all previous components. 
The generation process follows a four-stage pipeline: individual API call generation, relationship-aware sequencing, constraint satisfaction, and targeted mutation.

\smallskip \noindent \textbf{Stage 1: Individual API Call Generation.}
For each target API function, we use the corresponding CFG from the Grammar Generation component to generate syntactically valid API calls. 
The CFG rules guide the systematic construction of parameter values, ensuring type correctness and adherence to documented constraints. 
This stage produces a pool of well-formed individual API calls that serve as building blocks for sequence construction.

\smallskip \noindent \textbf{Stage 2: Relationship-Aware Sequencing.}
To construct API call sequences (denoted as \(c_1, c_2, ..., c_n\)), we begin with a seed API call \(c_1\) and consult the relationships identified by the Relationship Inference component. The sequencing strategy differs based on relationship tier:

\smallskip \noindent (a) For candidate relationships, we select related API functions and generate them independently using their respective CFGs. 
These calls are concatenated without parameter-level coupling, as candidate relationships only require co-occurrence without specific parameter constraints.

\smallskip \noindent (b) For strong symbolic relationships, we enforce parameter-level dependencies during generation. Starting with the parameter values generated for \(c_1\), we extract the symbolic constraints from the Relationship Inference component and instantiate them with concrete values. 
We then invoke an SMT solver (Z3) to determine parameter values for \(c_2\) that satisfy the relationship constraints. 
This process continues iteratively to build longer sequences that respect all symbolic dependencies. 
The maximum sequence length is configurable, and based on our empirical evaluation, we set it to 2,000 API calls. 
Longer sequences could result in excessive PDF execution time, making it difficult to distinguish between normal processing and potential hangs during testing.


\smallskip \noindent \textbf{Stage 3: Object-API Relationship Integration.}
After generating JavaScript API call sequences, \ourtool also needs PDF document structures that make those calls meaningful.
We therefore incorporate Cooper~\cite{DBLP:conf/ndss/cooper}, a cooperative mutation technique that links JavaScript APIs with PDF native objects, i.e., structural objects inside the PDF file such as pages, form fields, actions, and annotations.
The resulting object-API relationships specify which native object classes should appear in a generated document, and are separate from the inter-API relationships among JavaScript calls.
Cooper builds these relationships by clustering native objects from a PDF corpus and identifying object classes correlated with successful execution of API groups.
During test-case construction, after generating API call sequences, Cooper consults this object-API relationship map to identify relevant PDF object classes that should be present in the document. 
Cooper enriches our test cases by extracting PDF objects from sample documents and applying targeted mutations and combinations to these objects. 
This produces richer PDF documents with diverse native structures, extending coverage of PDF object processing code paths while ensuring the presence of native structures needed to support JavaScript operations.

\smallskip \noindent \textbf{Stage 4: Targeted Mutation. }
The previous three stages are designed to generate syntactically and semantically valid JavaScript code that adheres to API specifications and constraints extracted from official manuals, producing test cases that should execute without crashes under normal circumstances.
To test the robustness of the target JavaScript engines when receiving malformed inputs, we apply selective mutations to a subset (approximately 15\%) of the generated test cases. 
For strings, we target decoder and escape handling by injecting non-BMP Unicode, escape sequences (e.g., \texttt{\textbackslash uXXXX}, \texttt{\textbackslash xXX}), and malformed encodings.
For numbers, we target conversion edge cases using boundary values, \texttt{Infinity}, \texttt{NaN}, scientific-notation variants, and hexadecimal literals.

The complete generation pipeline produces test cases ranging from specification-compliant inputs (for functional testing) to deliberately malformed inputs (for security testing).
Our implementation extracts 435 API functions with 507 parameters from documented specifications and 231 API functions with 379 parameters from undocumented ones, yielding CFG-based rules for 886 parameters across 666 API functions.
%
Function selection follows a weighted random distribution based on relationship frequency.

\smallskip \noindent \textbf{Running Example.}
Stage~1 instantiates each API call invocation from its grammar, e.g., \func{app.popUpMenuEx({   cName:"popup", oSubMenu:\{label:"kM", action:"mFv"\}})} and \func{search.query({cQuery:?})}.
Stage~2 substitutes \func{x="popup"} into \texttt{(=\ x\ y)} and Z3 solves \func{y="popup"}, fixing \func{cQuery="popup"}.
Stage~3 (Cooper) supplies the necessary native PDF objects (e.g., the form-field/AcroForm and page objects that the API calls operate on).
Stage~4 may inject Unicode escapes via mutation.
The result is exactly Listing~\ref{lst:case2}, the test case that triggered a use-after-free vulnerability later fixed by Foxit and assigned a CVE entry.

\section{Evaluation Methodology}
In this section, we detail our evaluation methodology for \ourtool.
We compare \ourtool against current state-of-the-art techniques in two separate experiments: (1) we assess how much code can be reached (coverage), and (2) we test how many vulnerabilities can be found. 

\subsection{Experiment Setup}
We compare \ourtool against the current three state-of-the-art fuzzers for PDF readers: TypeOracle~\cite{DBLP:conf/icse/typeoracle}, Favocado~\cite{DBLP:conf/ndss/favocado}, and Cooper~\cite{DBLP:conf/ndss/cooper}.
These fuzzers start by generating a large input corpus for the target PDF readers and iteratively run each generated target through the PDF reader, observing its behavior.
This is similar to \ourtool's approach, making it easy to compare against these tools.

Since \ourtool leverages LLMs, we also compare our tool against two general-purpose LLM-based fuzzing techniques.
The first is Fuzz4All, which is an LLM-based universal fuzzer capable of generating test cases for various types of inputs, independent of the fuzzing target~\cite{DBLP:conf/icse/fuzz4all}.
It does so by using two LLMs: a distillation LLM and a generation LLM.
The distillation LLM takes arbitrary input related to the to-be-generated test cases (such as documentation or source code) and generates a prompt that will be used by the generation LLM to create fuzzing inputs.
As a last comparison, we use a naive LLM-based approach inspired by TitanFuzz~\cite{DBLP:conf/issta/titanfuzz}, which assumes that documentation related to the to-be-generated fuzzing input is already part of an LLM's training corpus.
Under these assumptions, there is no need to provide additional input related to the API manual since it is already part of the model's data.
To generate an input corpus, we therefore simply prompt the LLM to generate test cases for fuzzing the JavaScript engine of PDF fuzzers.

For targets, we selected three widely used PDF readers with built-in JavaScript engines: Adobe Acrobat Reader v24.005.20421, Foxit PDF Reader v2024.4.0.27683, and PDF-XChange Editor v10.5.2.395 (all of which were the latest versions at the time of this study).

\subsection{Experiment Design and Metrics}
We configure Cooper~\cite{DBLP:conf/ndss/cooper}, Favocado~\cite{DBLP:conf/ndss/favocado} using their default settings.
For TypeOracle~\cite{DBLP:conf/icse/typeoracle}, in addition to using its default settings, we also compare its integration with Cooper~\cite{DBLP:conf/ndss/cooper} and Favocado~\cite{DBLP:conf/ndss/favocado}.
For Fuzz4All~\cite{DBLP:conf/icse/fuzz4all}, we use the JavaScript API manual~\cite{adobe_acrobat_manual, adobe_acrobat_manual2} as input.
By default, Fuzz4All uses GPT-4 for the distillation LLM and StarCoder for the generation LLM.
However, both models are outdated, and newer versions are available.
As such, we run Fuzz4All using GPT-4o for the distillation LLM and StarCoder2 for the generation LLM.
For the naive vanilla-LLM method, we test three of the most popular contemporary models: GPT-4o~\cite{gpt-4o} (OpenAI's foundational general-purpose model), GPT-o3-mini~\cite{o3-mini} (OpenAI's reasoning model), and Claude-3.7-Sonnet~\cite{claude-3.7} (Anthropic's reasoning model), which were the latest versions available at the time of the experiment. 

Our evaluation does not compare different online fuzzing loops.
Instead, we standardize runtime exploration (fuzzing) across tools and compare only the different test-case generators.
Concretely, prior systems such as Cooper~\cite{DBLP:conf/ndss/cooper} and Favocado~\cite{DBLP:conf/ndss/favocado} primarily contribute test-case-generation components.
TypeOracle~\cite{DBLP:conf/icse/typeoracle} additionally provides an open-source wrapper for lightweight execution.
In our experiments, we use TypeOracle's public fuzzing wrapper uniformly for all methods.
The runtime process is therefore identical across tools: generate test cases, execute and monitor them, and record the outcomes.
It does not involve an AFL-style seed queue, power schedule, mutation strategy, or any coverage-guided feedback loop.
Accordingly, any substantive difference among the compared systems stems from the quality of the generated test cases, rather than any different scheduling or feedback mechanisms.

\noindent \textbf{Code Coverage.}
Because our targets are closed-sourced, we cannot directly instrument the PDF readers to collect code coverage.
Instead, we rely on DynamoRIO~\cite{DynamoRIO} to dynamically instrument each target program and collect basic block coverage.
For each tool, we perform five independent 24-hour runs on each target PDF reader.

\noindent \textbf{Vulnerability Discovery.}
Due to the significant execution overhead and instability introduced by dynamic instrumentation tools like DynamoRIO (which can cause false positive crashes), we separate our vulnerability detection experiments from coverage collection.
We let each tool generate an input corpus of test cases for each PDF reader.
For each input, we run it through all three targets and detect when a target crashes using the Windows-provided \texttt{werfault.exe}~\cite{WerFault}.
We then manually analyze each crash to determine whether it was caused by a vulnerability or not.
We ran each tool for two weeks on all of the target PDF readers.


Each experiment runs in a VMware Workstation virtual machine (VM) hosted on a system with an 8-core Intel Core i7-7700 processor (3.60GHz) and 64GB RAM.
Each VM is configured with 4 CPU cores, 8GB of RAM, and Windows 8.1 - chosen for its lower resource requirements compared to newer Windows versions. 
All of our target PDF readers are fully compatible with Windows 8.1, ensuring that their use does not adversely affect the results.
Whenever a crash was found by a tool, we verified its reproducibility in a newer version of Windows (Windows 10 22H2 and Windows 11 24H2) to ensure compatibility with newer operating systems.

\section{Evaluation Results}





\subsection{Basic Block Coverage}
\label{sec:eval_performance}

\begin{figure*}[t!] 
\centering
\begin{subfigure}{0.31\textwidth}
\includegraphics[width=\linewidth]{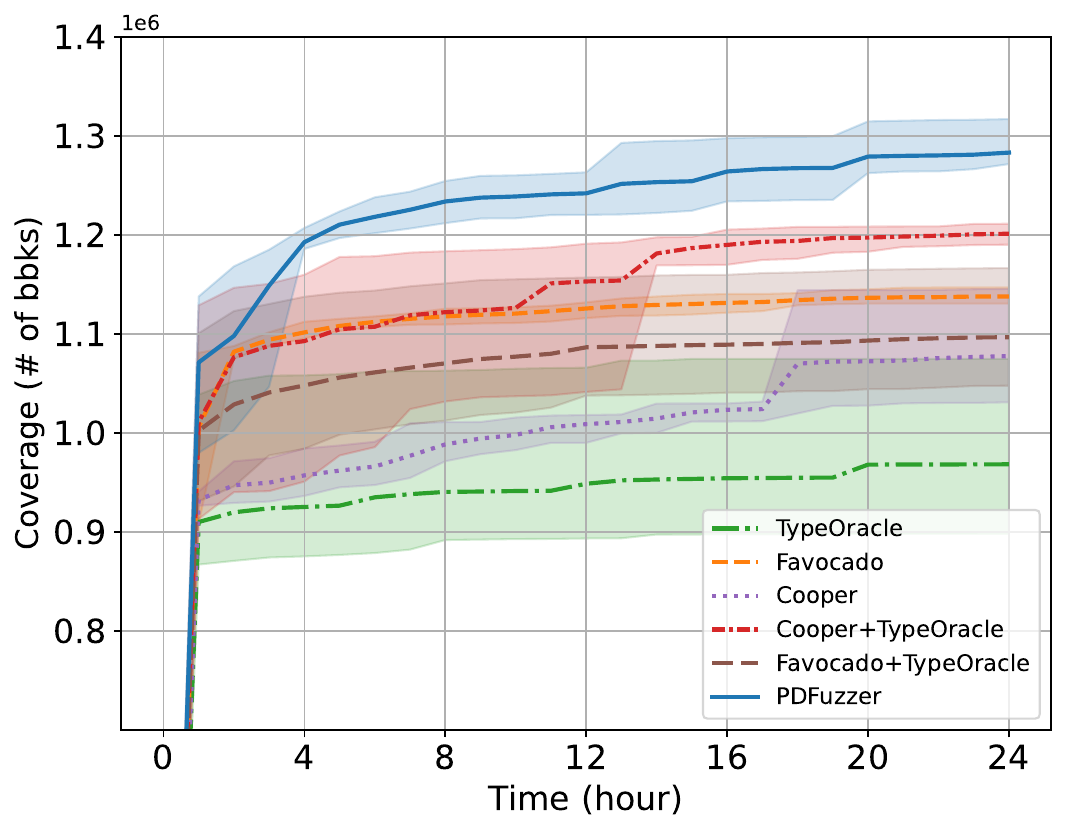}
\caption{\texttt{Adobe Acrobat Reader}} \label{fig:adobe_traditional}
\end{subfigure}
\begin{subfigure}{0.33\textwidth}
\includegraphics[width=\linewidth]{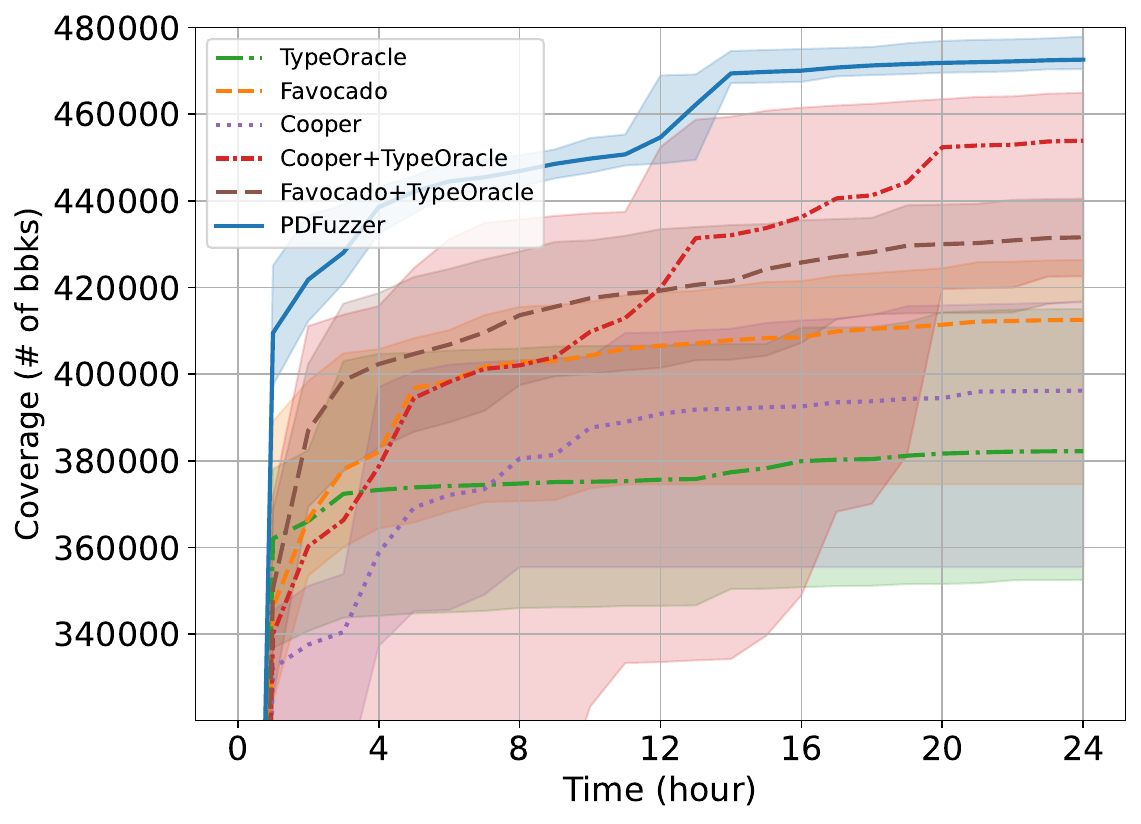}
\caption{\texttt{Foxit PDF Reader}} \label{fig:foxit_traditional}
\end{subfigure}
\begin{subfigure}{0.33\textwidth}
\includegraphics[width=\linewidth]{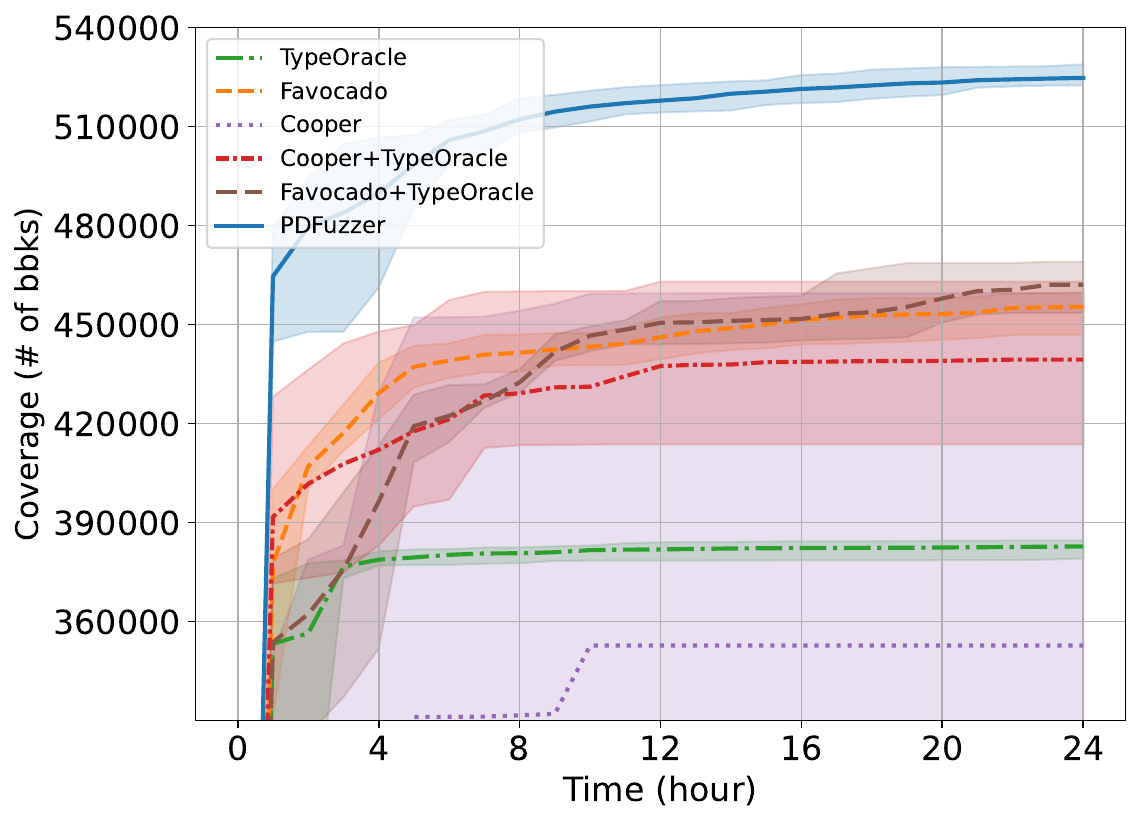}
\caption{\texttt{PDF-XChange Editor}} \label{fig:xchange_traditional}
\end{subfigure}
\begin{subfigure}{0.31\textwidth}
\includegraphics[width=\linewidth]{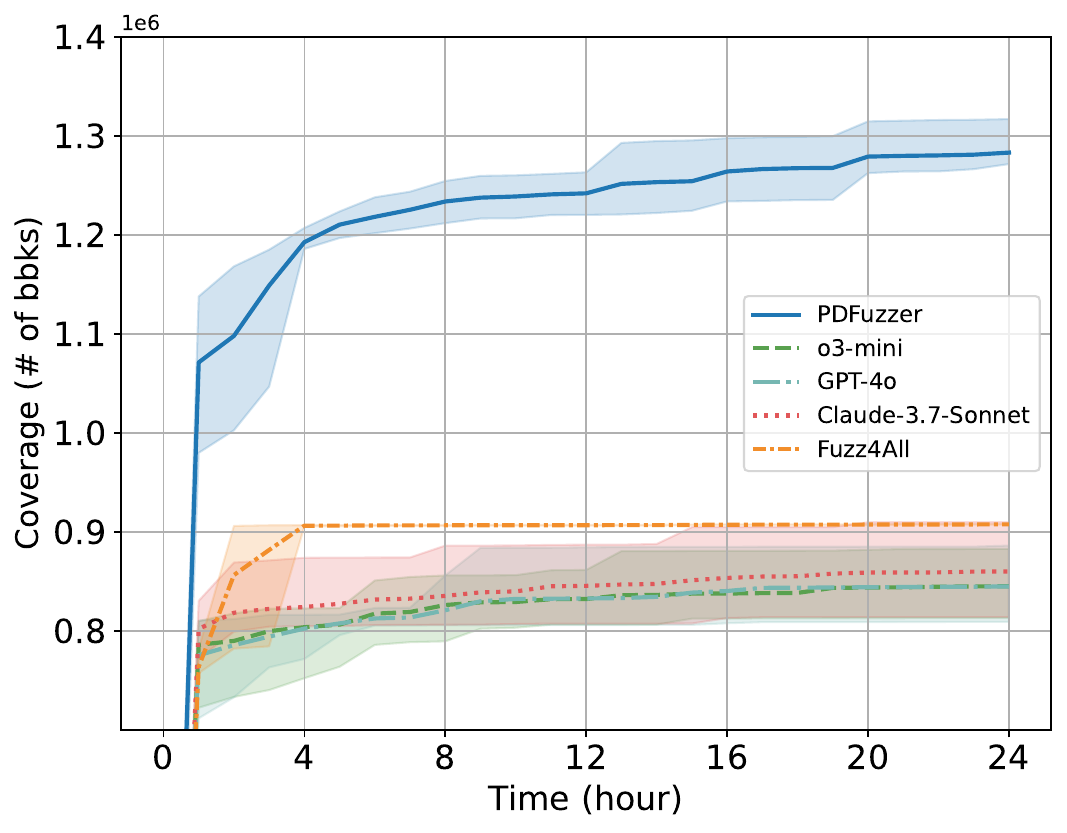}
\caption{\texttt{Adobe Acrobat Reader}} \label{fig:adobe_llm}
\end{subfigure}
\begin{subfigure}{0.33\textwidth}
\includegraphics[width=\linewidth]{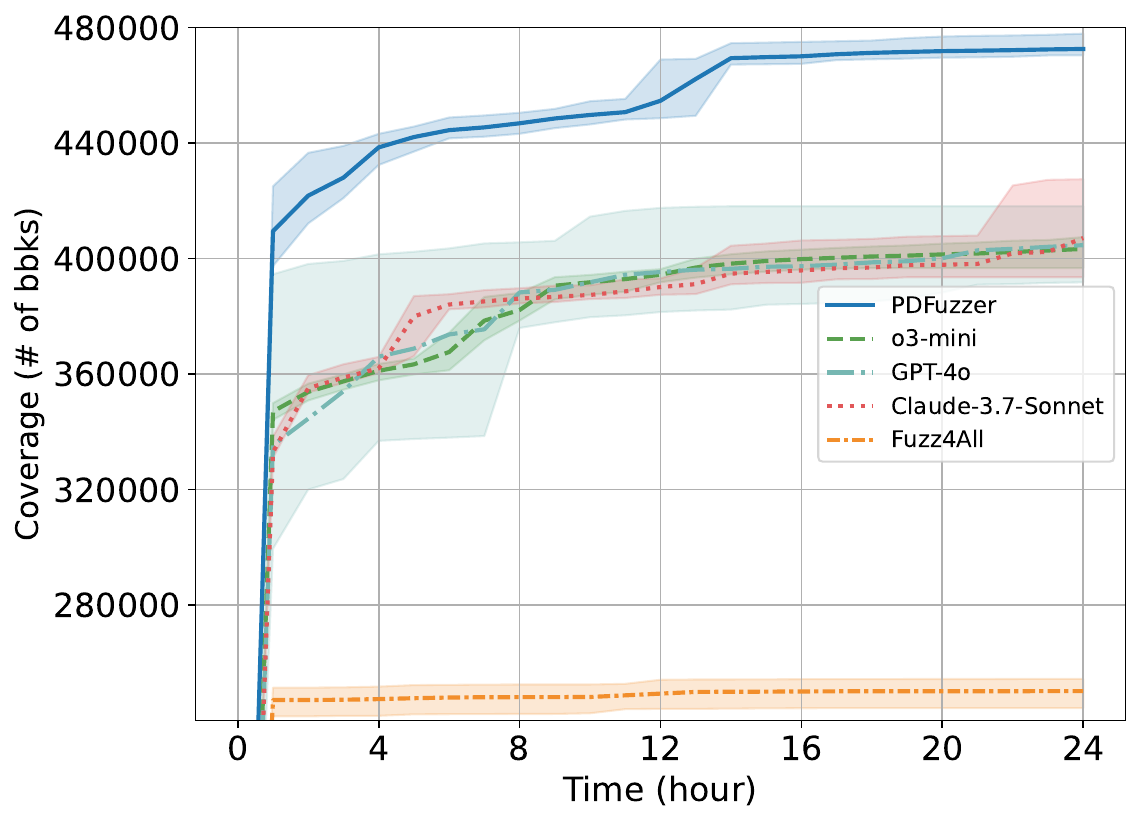}
\caption{\texttt{Foxit PDF Reader}} \label{fig:foxit_llm}
\end{subfigure}
\begin{subfigure}{0.33\textwidth}
\includegraphics[width=\linewidth]{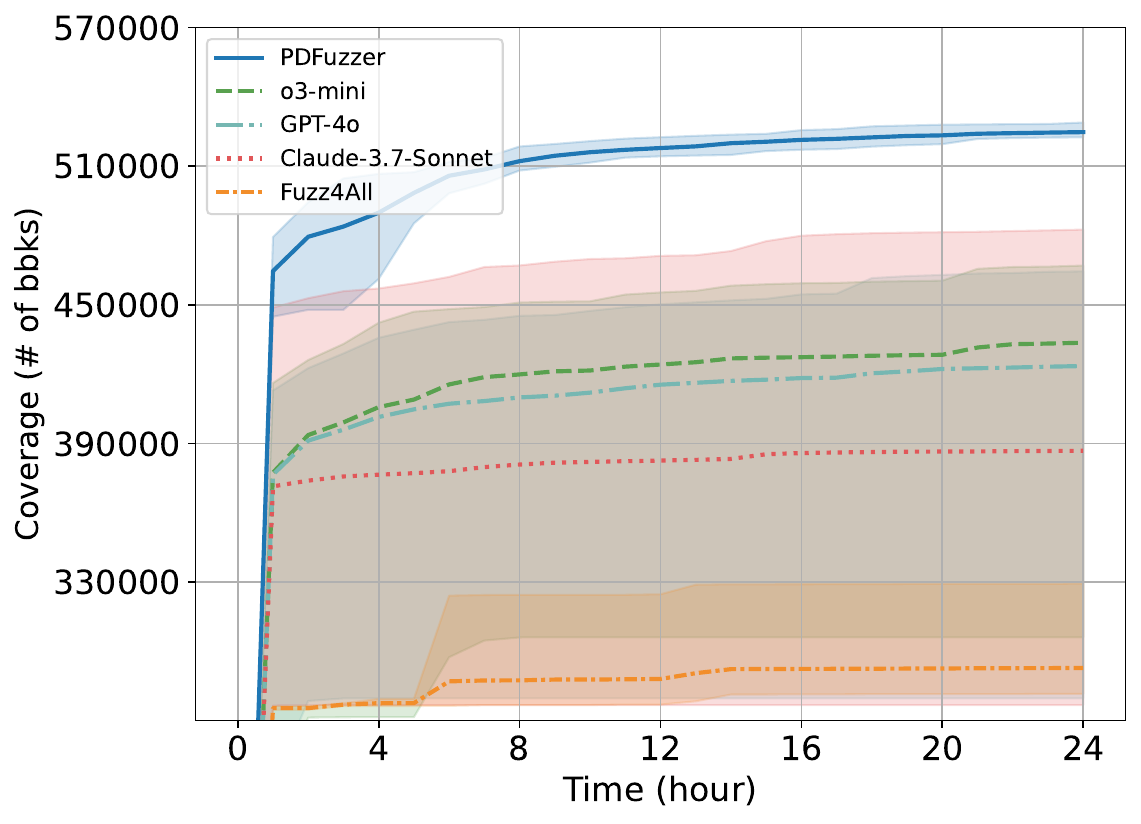}
\caption{\texttt{PDF-XChange Editor}} \label{fig:xchange_llm}
\end{subfigure}
\caption{(Top row) \ourtool vs. traditional tools. (Bottom row) \ourtool vs. LLM-based tools.}
 \label{fig:coverage_results_llm}
\end{figure*}
Figure~\ref{fig:coverage_results_llm} shows the mean basic-block coverage achieved by each tool across all three targets.
We observe that \ourtool (blue curve) consistently achieves the highest coverage, improving by up to 37\% over TypeOracle (up to 19\% over TypeOracle+Cooper and up to 17\% over TypeOracle+Favocado), up to 15\% over Favocado, and up to 48\% over Cooper.
These gains demonstrate that our higher-quality inputs indeed explore substantially more code than existing methods.

For the naive LLM-based approach, Claude-3.7-Sonnet generally performs best among LLMs, yet still underperforms \ourtool.
Specifically, \ourtool achieves between 16\% and 49\% higher coverage compared to Claude-3.7-Sonnet.
For Fuzz4All, we notice that the majority of generated test cases were invalid JavaScript code, mostly containing C code or Java code, or even just plain text.
As a result, Fuzz4All has the lowest coverage on both Foxit and PDF-XChange.
For Adobe, Fuzz4All is not the worst performer, which we assume is due to Adobe's more extensive error handling within the PDF reader.

Listing~\ref{lst:example1} shows an example input generated by \ourtool that demonstrates a value-constraint relationship between \func{customDictionaryCreate} and \func{addWord}, which only \ourtool captured in its input corpus.
The \func{customDictionaryCreate} call creates a custom dictionary with a unique identifier specified by the \func{cName} parameter. 
This operation sets up the necessary precondition for any subsequent operations that depend on an existing dictionary. 
In the next step, \func{addWord} adds a new word into this dictionary referencing the same \func{cName} value from the previous call.
This order of execution ensures that \func{addWord} is not executed without a prior successful call to \func{customDictionaryCreate}, maintaining control of the necessary dependency.
Furthermore, the generated test case includes a valid \func{cLanguage} value (\texttt{en_US}).
Although the type of the \func{cLanguage} parameter is \texttt{String}, an additional value constraint enforces that the string is a valid language code.
\ourtool was able to capture this constraint and enforce it in this test case, demonstrating that merely knowing the parameter type (as done by previous work) is typically insufficient to generate valid parameter values. 

\begin{lstlisting} [caption={Example that \ourtool can capture more relationships and generate higher quality arguments}\label{lst:example1},style=htmlcssjs]
spell.customDictionaryCreate({
    cName: "\\xfe\\xff\\u1a18\\ud428\\uf9cc", 
    bShow: false, 
    cLanguage: "en_US"})
spell.addWord({
    cName: "\\xfe\\xff\\u1a18\\ud428\\uf9cc"}),
    cWord: "\\uea83\\u5e92\\uda70\\udcd4\\uf619\\u1775\\ud827\\udce1"})
\end{lstlisting}

\begin{table*}[!t]
\centering
\caption{Discovered Zero-Day Vulnerabilities.
Severity: ACE (High Severity), Information Disclosure (Moderate Severity), DoS (Low Severity); 
T: TypeOracle, F$^1$: Favocado, L: Pure LLM, F$^2$: Fuzz4All, C: Cooper, F+T: Favocado+TypeOracle, C+T: Cooper+TypeOracle; P$^a$: \ourtool with Function-level Grammar, P$^p$: \ourtool with Param-level Grammar, P$^c$: \ourtool with Candidate Relation;  P$^s$: \ourtool with Strong Relation, P$^f$: \ourtool-Full}
\label{tab:vulnerabilities}
\adjustbox{max width=\textwidth}{%
\small

\begin{NiceTabular}{cllll@{\hspace{3pt}}c@{\hspace{3pt}}c@{\hspace{3pt}}c@{\hspace{3pt}}c@{\hspace{3pt}}c@{\hspace{3pt}}c@{\hspace{3pt}}c@{\hspace{5pt}}||c@{\hspace{3pt}}c@{\hspace{3pt}}c@{\hspace{3pt}}c@{\hspace{3pt}}c}
\CodeBefore
   \rowcolors{2}{gray!15}{}
\Body

\toprule
\multirow{2}{*}{\textbf{ID}} & \multirow{2}{*}{\textbf{Target}} & \multirow{2}{*}{\textbf{Type}} & \multirow{2}{*}{\textbf{Potential Impact}} & \multirow{2}{5em}{\textbf{Status}} & \multicolumn{7}{c}{\textbf{Baseline Tools}} & \multicolumn{5}{c}{\textbf{\ourtool Variants}} \\
\cmidrule(lr){6-17}
& & & & & \textbf{T} & \textbf{F$^1$} & \textbf{L} & \textbf{F$^2$} & \textbf{C} & \textbf{F+T} & \textbf{C+T} & \textbf{P$^a$} & \textbf{P$^p$} & \textbf{P$^c$} & \textbf{P$^s$} & \textbf{P$^f$} \\
\midrule
1 & Adobe & Out-of-bounds Write & Arbitrary Code Execution & CVE-2025-43575 & & & & & & & & & & & $\checkmark$ & $\checkmark$ \\
 2 & Adobe & Memory Corruption & Arbitrary Code Execution & Confirmed & $\checkmark$ & $\checkmark$ & & & $\checkmark$ & $\checkmark$ & $\checkmark$ & & $\checkmark$ & $\checkmark$ & $\checkmark$ & $\checkmark$ \\
3 &Adobe & Buffer Overflow & Arbitrary Code Execution & Fixed & $\checkmark$ & $\checkmark$ & & & & $\checkmark$ & $\checkmark$ & & & $\checkmark$ & $\checkmark$ & $\checkmark$ \\
 4 & Adobe & Use-after-free & Arbitrary Code Execution & Fixed & & & & & & & & & & & & $\checkmark$ \\
5 & Foxit & Use-after-free & Arbitrary Code Execution & CVE-2026-3777 & & & & & & & & & & & $\checkmark$ & $\checkmark$ \\
 6 & Foxit & Use-after-free & Information Disclosure & CVE-2025-55308 & & $\checkmark$ & & & & $\checkmark$ & $\checkmark$ & & & $\checkmark$ & $\checkmark$ & $\checkmark$ \\
7 & Foxit & Use-after-free & Arbitrary Code Execution & CVE-2025-55314 & & & & & & & & & & & $\checkmark$ & $\checkmark$ \\
 8 & Foxit & Buffer Overflow & Arbitrary Code Execution & CVE-2025-55312 & & & & & & & & & & & $\checkmark$ & $\checkmark$ \\
9 & Foxit & Out-of-bounds Read & Information Disclosure & Fixed & $\checkmark$ & & & & & $\checkmark$ & $\checkmark$ & $\checkmark$ & $\checkmark$ & $\checkmark$ & $\checkmark$ & $\checkmark$ \\
 ... & Foxit & Out-of-bounds Read & Information Disclosure & Fixed & & & & & & & & & & & $\checkmark$ & $\checkmark$ \\
15 & Foxit & Out-of-bounds Read & Information Disclosure & CVE-2025-55307 & & & & & & & & & & $\checkmark$ & $\checkmark$ & $\checkmark$ \\
 16 & Foxit & Out-of-bounds Read & Information Disclosure & Fixed & & & & & & & & & $\checkmark$ & $\checkmark$ & $\checkmark$ & $\checkmark$ \\
17 & Foxit & Out-of-bounds Read & Information Disclosure & Fixed & & & & & & & & & & $\checkmark$ & $\checkmark$ & $\checkmark$ \\
 18 & Foxit & Uncontrolled Recursion & Denial-of-service & CVE-2026-3778 & & & & & & & & & & $\checkmark$ & $\checkmark$ & $\checkmark$ \\
19 & Foxit & Null-pointer-dereference & Denial-of-service & CVE-2025-55313 & & & & & & & & & & & $\checkmark$ & $\checkmark$ \\
 20 & Foxit & Null-pointer-dereference & Denial-of-service & CVE-2026-3776 & & & & & & & & $\checkmark$ & $\checkmark$ & $\checkmark$ & $\checkmark$ & $\checkmark$ \\
21 & Xchange & Memory Corruption & Arbitrary Code Execution & CVE-2025-6661 & & & & & & & & & & & $\checkmark$ & $\checkmark$ \\
 22 & Xchange & Use-after-free & Arbitrary Code Execution & Fixed & & & & & & & & & & & $\checkmark$ & $\checkmark$ \\
23 & Xchange & Use-after-free & Arbitrary Code Execution & Fixed & & & & & & & & & & & $\checkmark$ & $\checkmark$ \\
 24 & Xchange & Use-after-free & Arbitrary Code Execution & Fixed & & & & & & & & & & & & $\checkmark$ \\
25 & Xchange & Out-of-bounds Read & Information Disclosure & Fixed & & & & & & & & & $\checkmark$ & $\checkmark$ & $\checkmark$ & $\checkmark$ \\
 ... & Xchange & Null-pointer-dereference & Denial-of-service & Submitted & & & & & & & & & & & $\checkmark$ & $\checkmark$ \\
29 & Xchange & Null-pointer-dereference & Denial-of-service & Submitted & $\checkmark$ & $\checkmark$ & $\checkmark$ & & & $\checkmark$ & $\checkmark$ & $\checkmark$ & $\checkmark$ & $\checkmark$ & $\checkmark$ & $\checkmark$ \\
 30 & Xchange & Stack Exhaustion & Denial-of-service & Fixed & $\checkmark$ & & & & & $\checkmark$ & $\checkmark$ & $\checkmark$ & $\checkmark$ & $\checkmark$ & $\checkmark$ & $\checkmark$ \\
31 & Xchange & Stack Exhaustion & Denial-of-service & Submitted & & & & & $\checkmark$ & & & & & & & $\checkmark$ \\

\multicolumn{4}{l}{Aggregated Results}  & & 5 & 4 & 1 & 0 & 2 & 6 & 6 & 4 & 7 & 12 & 28 & 31 \\
\bottomrule
\end{NiceTabular}}
\end{table*}

\subsection{Vulnerability Discovery}
\label{sec:eval_vulnerability}

During the two-week vulnerability discovery campaign, \ourtool generated 57 test cases that consistently triggered crashes across the three PDF readers (23 in Adobe Acrobat Reader, 24 in Foxit PDF Reader, and 10 in PDF-XChange Editor). 
We manually triaged these inputs in two steps.
First, for each test case, we performed statement-level reduction by iteratively deleting JavaScript statements and keeping the smallest subset that still reproduces the crash. 
Second, we deduplicated the minimized crashes by comparing normalized call-stack signatures, following TypeOracle~\cite{DBLP:conf/icse/typeoracle} and Cooper~\cite{DBLP:conf/ndss/cooper}. 
After triage, the 57 crashing inputs collapse to 31 unique crash signatures, which we report as 31 distinct zero-day vulnerabilities discovered by \ourtool.
None of the other tools was able to find a vulnerability that \ourtool was not able to find.
We reported all vulnerabilities to the respective parties; two of the vulnerabilities have already received a bounty for a total of \$2,450.

Table~\ref{tab:vulnerabilities} lists all vulnerabilities found in the three PDF reader targets.
Of the 31 unique zero-day vulnerabilities we discovered, 11 are high-severity issues that can potentially lead to arbitrary code execution (ACE). 
As of this writing, vendors have fixed 26 of the 31 vulnerabilities, and 10 have been assigned CVE entries.
Below, we discuss one case study of the vulnerabilities found by \ourtool.
 

\smallskip \noindent \textbf{Case Study: Vulnerability Involving Strong Relationship.}
The running example introduced in Section~\ref{sec:running_example} (Listing~\ref{lst:case2}) is precisely the test case generated by \ourtool that triggered this use-after-free vulnerability in Foxit PDF Reader, corresponding to ID~5 in Table~\ref{tab:vulnerabilities} (later fixed and assigned a CVE entry).
Specifically, the first API function \func{app.popUpMenuEx}, which creates a pop-up menu at the current mouse position, allocates a memory region through dynamic memory management.
The second API function \func{search.query} searches for characters corresponding to \func{cQuery} within the document scope, but the call to \func{search.query} releases the memory region allocated by \func{app.popUpMenuEx}.
When we call the popup menu-related API functions again, this memory address is accessed again, causing a use-after-free vulnerability.
As can be seen, this vulnerability requires a value-constraint relationship between the \func{cName} parameter of \func{app.popUpMenuEx} and the \func{cQuery} parameter of \func{search.query}.
It is extremely unlikely to generate a test case where these two values are equal through TypeOracle's~\cite{DBLP:conf/icse/typeoracle} random generation.
This relationship also goes beyond Favocado's~\cite{DBLP:conf/ndss/favocado} type-based matching between return value and parameter.
Therefore, it can only be discovered by \ourtool.
In our two-week evaluation, this bug was only triggered when \ourtool enabled strong symbolic relationships ($P^s/P^f$ in Table~\ref{tab:vulnerabilities}), while baseline tools (TypeOracle, Favocado, Cooper, and their combinations) did not trigger it.



\subsection{Inferred Relationship Distribution}
\label{sec:eval_relation_breakdown}

Among 1,019 inferred strong symbolic relationships, 887 (87.0\%) are value-constraint, 63 (6.2\%) are implicit, and only 69 (6.8\%) are producer-consumer; a producer-consumer-only model would miss most of the strong symbolic relationships \ourtool instantiates.
Table~\ref{tab:relation_distribution} additionally breaks down value-constraint subtypes (showing patterns beyond direct equality).
\begin{table}[t]
\caption{Distribution of inferred strong symbolic relationships, including subtypes of value-constraint relationships.}
\label{tab:relation_distribution}
\small
\centering
\begin{adjustbox}{max width=\columnwidth}
\begin{tabular}{lrr}
\toprule
\textbf{Category} & \textbf{Count} & \textbf{Proportion} \\
\midrule
\multicolumn{3}{c}{\textbf{Strong symbolic relationships}} \\
\midrule
Producer-Consumer & 69 & 6.8\% \\
Value-Constraint & 887 & 87.0\% \\
Implicit & 63 & 6.2\% \\
\textbf{Total} & \textbf{1,019} & \textbf{100.0\%} \\
\midrule
\multicolumn{3}{c}{\textbf{Value-Constraint subtypes}} \\
\midrule
Direct Equality & 667 & 75.2\% \\
Relational Comparisons & 69 & 7.8\% \\
Set Memberships & 59 & 6.7\% \\
Range Constraints & 26 & 2.9\% \\
String Constraints & 8 & 0.9\% \\
Complex Relationships & 58 & 6.5\% \\
\textbf{Total} & \textbf{887} & \textbf{100.0\%} \\
\bottomrule
\end{tabular}
\end{adjustbox}
\end{table}

\subsection{Strong Symbolic Relationship Instantiation}
\label{sec:eval_instantiation}

A strong symbolic relationship is useful only if its constraints can be solved during generation to produce a concrete API call sequence.
To assess how reliably this happens, we generated 1,000 PDF test cases and tracked, for every selected strong symbolic relationship, whether the Test Case Generator (Section~\ref{sec:method}, Stage~2) successfully produced a satisfying assignment.
Table~\ref{tab:relation_instantiation} reports the resulting success rates, broken down by semantic type.
\begin{table}[t]
\caption{Generation success and failure breakdown by semantic type for strong symbolic relationships over 1,000 generated PDF test cases.}
\label{tab:relation_instantiation}
\small
\centering
\begin{adjustbox}{max width=\columnwidth}
\begin{tabular}{lrrrr}
\toprule
\textbf{Type} & \textbf{Invocations} & \textbf{Success} & \textbf{Failure} & \textbf{Rate} \\
\midrule
Producer-Consumer & 4,704 & 3,832 & 872 & 81.5\% \\
Value-Constraint & 60,675 & 57,706 & 2,969 & 95.1\% \\
Implicit & 6,207 & 3,970 & 2,237 & 64.0\% \\
\textbf{Total} & \textbf{71,586} & \textbf{65,508} & \textbf{6,078} & \textbf{91.5\%} \\
\bottomrule
\end{tabular}
\end{adjustbox}
\end{table}

Out of 71,586 selected strong symbolic relationships, 65,508 (91.5\%) are successfully instantiated, indicating that the relationships \ourtool infers are practical to enforce during generation.
Value-constraint relationships achieve the highest success rate (95.1\%) because their constraints are typically simple equalities or set-membership conditions that Z3 solves directly; producer-consumer relationships reach 81.5\%, occasionally failing when an upstream parameter has already been bound to an incompatible concrete value.
Implicit relationships are the hardest to instantiate (64.0\%), since they often require both a parameter constraint (e.g., a specific event name) and a state precondition (e.g., the target field being focusable).
Most of the remaining failures stem from two sources: (1) the targeted mutation pass (Stage~4) intentionally violates the constraint to probe robustness, and (2) implementation edge cases such as parameter values being assembled before the SMT solver can re-bind them; genuinely unsatisfiable constraint sets are rare.

\subsection{Cost Analysis}
To run \ourtool end-to-end, the total cost inferred by the LLM components is \$60.77, \$73.38, and \$92.36 for GPT-4o, o3-mini, and Claude, respectively.
For our naive LLM approach, this cost is significantly higher, rising to \$76.14, \$156.76, and \$174.83 for GPT-4o, o3-mini, and Claude, respectively.
\ourtool is, therefore, not only better at exploring code and finding vulnerabilities compared to the naive LLM method, but it also does so at a lower cost.
This makes our tool more accessible to be used in practice by users with limited resources.
Furthermore, the average generation time per test case for the naive LLM method exceeds 60 seconds for the OpenAI models and up to 90 seconds for Claude, highlighting both computational inefficiency and scalability limitations of pure LLM-based approaches.
In contrast, \ourtool generates approximately three outputs per second for strong symbolic relationships and can achieve speeds as high as 20 outputs per second for regular test cases.
This stark difference in throughput underscores the efficiency and scalability of \ourtool compared to conventional LLM-based methods.

\subsection{The Accuracy of LLM in Each Step}
\label{sec:llm_accuracy}

We randomly sampled 60 LLM outputs from each of \ourtool’s LLM-based steps and compared them to our manual annotations.
The sample size of 60 was chosen considering the manual annotation burden, where each output requires expert verification to determine correctness.

\smallskip \noindent \textbf{Specification Inference.} We reverse-engineered 60 undocumented API functions (covering 87 parameters total) to obtain their ground-truth type information. \ourtool achieved 94\% accuracy (82/87 correct) in identifying the parameter types, with mistakes spread across 4 API functions. 
Analysis of these errors reveals two distinct patterns in LLM reasoning. 
In successful cases, the LLM effectively corrected TypeOracle's misclassifications by leveraging semantic understanding from parameter names. 
For example, when TypeOracle incorrectly classified the \func{cFileFilter} parameter of \func{browseForMultipleDocs} as a Boolean, the LLM correctly identified it as a String type based on the "Filter" naming convention and browsing context. 
However, the LLM exhibited systematic errors when parameter names strongly suggested Boolean semantics but actual implementations required string types.
In the \func{dcSignup} API function, both \func{cEmailPerm} and \func{cConnectPerm} parameters were incorrectly classified as Boolean due to the ``Perm'' suffix implying permissions, when they actually accept string values representing permission levels. 


\smallskip \noindent \textbf{Grammar Generation.} On 60 API functions (covering 30 documented functions with 45 parameters and 30 undocumented functions with 58 parameters), \ourtool achieves 97\% accuracy (58/60 correct) for APIs and 98\% for parameters (101/103 correct).


\smallskip \noindent \textbf{Relationships Inference.} On 60 relation pairs, \ourtool achieved 93\% accuracy (56/60 correct), with three errors in API sequence identification and one in constraint inference.


The high accuracy of LLMs across the pipeline underscores our effective design, while \ourtool's overall performance suggests that occasional LLM errors have minimal impact.
It is worth noting that, as a fuzzing method, our seed generation accuracy is already remarkably high compared to traditional fuzzers that rely on random mutations, where valid inputs are far rarer.
Moreover, accuracy is not the typical primary goal in fuzzing; instead, the emphasis lies in broad coverage and exposing edge cases.
Imperfect inputs from LLM errors may even be beneficial by contributing to the exploration of unexpected program behaviors, aligning with fuzzing's objective of uncovering vulnerabilities through diverse inputs.

\section{Ablation Experiments}
\label{sec:ablation}

We validate the efficacy of \ourtool's components along four dimensions:
(1) undocumented API specification inference;
(2) context-free grammar generator;
(3) relationship symbolizing, and
(4) PDF objects generation integration.
Additionally, we evaluate the robustness of \ourtool’s performance across different LLMs.
Results are shown in Table~\ref{table:component_ablation}.

\begin{table}[t]
\caption{Component ablation study along four dimensions.}

\small
\centering
\begin{adjustbox}{max width=\columnwidth}
\begin{tabular}
{c|c|c||r|l}

\toprule

\textbf{Dims.} & \textbf{Target} & \textbf{Setting} & \textbf{24H Cov.} & \textbf{Change} \\
\hline

\multirow{6}{4em}{\textbf{Dim. 1}:\\Undoc.\\APIs} 

& Adobe & \textbf{\ourtool} & 1008279 & -- \\
& Adobe & TypeOracle & 725776 & {\color{red} $\downarrow$ 28\%} \\
\cline{2-5}

& Foxit & \textbf{\ourtool} & 349635 & -- \\
& Foxit & TypeOracle & 279361 & {\color{red} $\downarrow$ 20\%} \\
\cline{2-5}

& XChange & \textbf{\ourtool} & 336724 & -- \\
& XChange & TypeOracle & 333408 & {\color{red} $\downarrow$ 1\%} \\

\hline

\multirow{6}{4em}{\textbf{Dim. 2}: Grammar Granularity} 

& Adobe & \textbf{Param-Level} & 1121910 & -- \\
& Adobe & Function-Level & 921160 & {\color{red} $\downarrow$ 18\%} \\

\cline{2-5}

& Foxit & \textbf{Param-Level} & 406448 & -- \\
& Foxit & Function-Level & 371901 & {\color{red} $\downarrow$ 9\%} \\

\cline{2-5}

& XChange & \textbf{Param-Level} & 440480 & -- \\
& XChange & Function-Level & 416297 & {\color{red} $\downarrow$ 5\%} \\

\hline

\multirow{15}{4em}{\textbf{Dim. 3}: Relation\\Tier}

& Adobe & \textbf{Strong} & 1183295 & -- \\
& Adobe & Candidate & 1141773 & {\color{red} $\downarrow$ 4\%} \\
& Adobe & PC-Only & 1133093 & {\color{red} $\downarrow$ 4\%} \\
& Adobe & Favocado & 1137989 & {\color{red} $\downarrow$ 4\%} \\
& Adobe & None & 1121910 & {\color{red} $\downarrow$ 5\%} \\
\cline{2-5}

& Foxit & \textbf{Strong} & 459634 & -- \\
& Foxit & Candidate & 425201 & {\color{red} $\downarrow$ 8\%} \\
& Foxit & PC-Only & 397916 & {\color{red} $\downarrow$ 13\%} \\
& Foxit & Favocado & 412543 & {\color{red} $\downarrow$ 10\%} \\
& Foxit & None & 406448 & {\color{red} $\downarrow$ 12\%}  \\
\cline{2-5}

& XChange & \textbf{Strong} & 500579 & --  \\
& XChange & Candidate & 466248 & {\color{red} $\downarrow$ 7\%}  \\
& XChange & PC-Only & 488575 & {\color{red} $\downarrow$ 2\%}  \\
& XChange & Favocado & 457543 & {\color{red} $\downarrow$ 9\%}  \\
& XChange & None & 440480 & {\color{red} $\downarrow$ 12\%}  \\

\hline

\multirow{6}{4em}{\textbf{Dim. 4}:\\PDF \\Objects} 

& Adobe & \textbf{\ourtool-full} & 1283296 & -- \\
& Adobe & \ourtool-js & 1183295 & {\color{red} $\downarrow$ 8\%} \\
\cline{2-5}

& Foxit & \textbf{\ourtool-full} & 472620 & -- \\
& Foxit & \ourtool-js & 459634 & {\color{red} $\downarrow$ 3\%} \\
\cline{2-5}

& XChange & \textbf{\ourtool-full} & 524782 & -- \\
& XChange & \ourtool-js & 500579 & {\color{red} $\downarrow$ 5\%} \\




\bottomrule
\end{tabular}
\end{adjustbox}
\label{table:component_ablation}
\end{table}

%


\vspace{0.01cm} \noindent \textbf{Dimension 1: Undocumented specification inference.} 
We compare the coverage of TypeOracle (which relies on execution traces to infer parameter types) and \ourtool when considering specification inference only for undocumented APIs.
To isolate the effect of specification inference, we disabled relationship inference in both approaches.
\ourtool achieves 28\% and 20\% higher coverage than TypeOracle on Adobe Acrobat Reader and Foxit PDF Reader, respectively.
For XChange Editor, where the gain is limited (1\%), our reverse-engineering showed that it does not implement any undocumented APIs, including those found in Adobe.

\vspace{0.01cm} \noindent \textbf{Dimension 2: Grammar Granularity.} 
We run \ourtool using only function-level grammars versus parameter-level grammars and compare the coverage.
Parameter-level grammars achieve 5--18\% higher coverage, as the LLM captures more detail when focusing on individual parameter descriptions rather than processing them all at once.


\vspace{0.01cm} \noindent \textbf{Dimension 3: Relationship Tiers and Semantic Types.}

We compare coverage across five relationship-modeling settings: \emph{None} (no inter-API relationships), \emph{Candidate} (candidate co-occurrence only), \emph{PC-Only} (\ourtool configured to use only producer-consumer strong symbolic relationships), \emph{Favocado}~\cite{DBLP:conf/ndss/favocado} (an existing fuzzer included as an external producer-consumer baseline), and \emph{Strong} (all three semantic types of strong symbolic relationships, i.e., producer-consumer, value-constraint, and implicit).
As shown in the Relation Tier rows of Table~\ref{table:component_ablation}, stronger relationship modeling consistently improves coverage.
PC-Only performs close to Favocado, confirming that our PC-Only configuration replicates Favocado-style producer-consumer matching. 
In contrast, Strong outperforms both PC-Only and Favocado on all targets, with a maximum gain of 15.5\% on Foxit PDF Reader.
This confirms that value-constraint and implicit relationships contribute beyond producer-consumer relationships.

\vspace{0.01cm} \noindent \textbf{Dimension 4: PDF objects generation integration.}
We evaluate the impact of integrating PDF objects generation with JavaScript API calls by comparing \ourtool-js (JavaScript only) against\\ \ourtool-full (with PDF objects integration via Cooper~\cite{DBLP:conf/ndss/cooper}). \\
\ourtool-full achieves 3--8\% higher coverage, validating the effectiveness of Stage 3 in Test Case Generation, where Cooper's cooperative mutation approach leverages relationships between PDF native objects and JavaScript APIs.
Since many vulnerabilities depend on specific PDF document structures alongside JavaScript code, \ourtool-full ensures that generated documents include the native structures necessary to support JavaScript operations. 


\vspace{0.01cm} \noindent \textbf{LLM Influence.} 
Our earlier experiments used GPT-4o (OpenAI, May 2024).
Here, we additionally evaluate Claude-3.7-Sonnet (Anthropic, February 2025) and o3-mini (OpenAI, February 2025, optimized for reasoning tasks), focusing our LLM ablation on Adobe Acrobat Reader.
We found that the choice of LLM has minimal impact on \ourtool.
Comparing test coverage with o3-mini, Claude-3.7-Sonnet, and GPT-4o, the differences remain within 10,000 basic blocks, less than 1\% of the total in Adobe Acrobat Reader.
Thus, using newer LLMs beyond GPT-4o does not substantially affect the overall performance or stability of \ourtool.

\section{Generalization Beyond PDF Readers}


While our evaluation primarily focuses on JavaScript engines in PDF readers, the fundamental methodology underlying \ourtool is not limited to this specific domain.
\ourtool takes API documentation as input, extracts machine-readable specifications, and then derives grammars and cross-API constraints for generating high-quality test inputs. 
Since rich API documentation is widely available across software ecosystems, the same pipeline can be applied beyond JavaScript engines in PDF readers.

To validate this generalization in a different setting, we applied our specification extraction and grammar construction pipeline to Microsoft Word’s VBA (Visual Basic for Applications) APIs. 
VBA is the macro language for automating Microsoft Office applications (e.g., Word, Excel, PowerPoint), which are widely deployed, closed-source Windows programs with large and distinct API surfaces. 
From the official documentation (5,920 pages), we extracted specifications for 2,925 APIs covering 3,768 parameters, and generated context-free grammars for all of them. 
We observed that the three relationship types defined in Section~\ref{sec:relation_definition} (producer–consumer, value-constraint, and implicit/state-based relationships) also occur frequently in VBA APIs.
This suggests that our relationship modeling generalizes beyond JavaScript in PDF readers.

Given these grammars and inferred relationships, the next step is straightforward: generate VBA macro test cases and embed them into Word documents for fuzzing—analogous to how \ourtool embeds JavaScript into PDF files. 
This feasibility study suggests that \ourtool can be readily adapted to other documentation-rich domains with large API surfaces.

\section{Related Work}
\label{sec:rw}

\vspace{0.01cm} \noindent \textbf{API Security.} 
Existing work on API security involves vulnerability detection through static analysis or dynamic testing, as well as methods to infer relationships and dependencies among APIs. 
These techniques have been applied to various targets, such as RESTful APIs~\cite{DBLP:conf/icse/restler, DBLP:conf/sigsoft/restfuzz2, DBLP:conf/uss/minerrest}, the Linux kernel~\cite{DBLP:conf/icse/HaoZLDQS22, DBLP:conf/ccs/difuse, DBLP:conf/sosp/healer, DBLP:conf/ccs/imf, DBLP:conf/usenix/ksg, DBLP:conf/ndss/mock, DBLP:conf/uss/moonshine, DBLP:conf/sp/syzdescribe, DBLP:conf/ccs/syzgen, DBLP:conf/sp/syzgen++, DBLP:conf/uss/Fleischer0BBLPK23}, JavaScript engines~\cite{DBLP:conf/ndss/favocado, DBLP:conf/ndss/cooper, DBLP:conf/icse/typeoracle}, open-source projects~\cite{DBLP:conf/sigsoft/GuZZK16, DBLP:conf/issta/LeL18}, smart contracts~\cite{DBLP:conf/icse/syntest}, microservices~\cite{DBLP:conf/sigsoft/Guo0WLJDXS20}, software libraries~\cite{DBLP:conf/kbse/NguyenPSNYWRN22, DBLP:conf/icse/ZhongMLJ20, DBLP:conf/ndss/apigraph, DBLP:conf/uss/fuzzgen, DBLP:conf/icse/graphfuzz}, and deep learning libraries~\cite{DBLP:conf/issta/docter}.
Static methods like SyzDescribe~\cite{DBLP:conf/sp/syzdescribe} and GraphFuzz~\cite{DBLP:conf/icse/graphfuzz} extract signatures from source code, while dynamic approaches use execution traces~\cite{DBLP:conf/icse/typeoracle} or symbolic execution~\cite{DBLP:conf/ccs/syzgen, DBLP:conf/sp/syzgen++}. 
However, closed-source PDF readers preclude direct application of these techniques, and their natural-language API manuals resist automated extraction.
For API relationship inference, static approaches like RESTler~\cite{DBLP:conf/icse/restler} identify producer-consumer dependencies through type matching, while dynamic methods like HEALER~\cite{DBLP:conf/sosp/healer} and APIGraph~\cite{DBLP:conf/ndss/apigraph} observe execution feedback. 
Recent work also applied neural networks for sequence prediction~\cite{DBLP:conf/issta/docter, DBLP:conf/sigsoft/GuZZK16,DBLP:conf/ndss/mock,DBLP:conf/issta/LeL18}.
Our approach differs by using LLMs to generate API rules and infer relationships, reducing manual effort and improving scalability. Unlike prior work that mostly relies on name-based matching, we model richer interactions, including value constraints and implicit dependencies.

\vspace{0.02cm} \noindent \textbf{LLM-based Fuzzing.} 
Prior work has explored using LLMs to facilitate other fuzzing targets, such as protocols~\cite{DBLP:conf/ndss/chatafl, DBLP:conf/sp/llmif, mgptfuzz}, the Linux kernel~\cite{kernelgpt}, interpreters~\cite{DBLP:conf/icse/fuzz4all}, deep learning libraries~\cite{DBLP:conf/issta/titanfuzz, DBLP:conf/icse/fuzzgpt}, command-line applications~\cite{prophetfuzz}, and open-source libraries~\cite{prompt_fuzz}. 
They use LLMs to generate or mutate testing cases.
For example, LLMs have been used to produce initial seeds~\cite{DBLP:conf/uss/busyboxllm, DBLP:conf/issta/titanfuzz, DBLP:conf/ndss/chatafl} or concrete test cases~\cite{DBLP:journals/tse/SchaferNET24, DBLP:conf/icse/fuzzgpt}, generate scripts for test case creation~\cite{DBLP:journals/corr/g2fuzz}, generate fuzz drivers~\cite{prompt_fuzz}, summarize documentation~\cite{DBLP:conf/icse/fuzz4all, DBLP:conf/sp/llmif}, and analyze source code to derive specifications~\cite{kernelgpt, prompt_fuzz}. 
In mutation, they can operate at either the byte level~\cite{DBLP:journals/corr/llamafuzz}, the token level~\cite{DBLP:conf/issta/covrl, DBLP:conf/issta/titanfuzz}, or the whole input level~\cite{DBLP:conf/icse/fuzz4all}.
Our experiments show that using LLMs to directly generate test cases for PDF reader fuzzing is slow and costly. 
Moreover, the closed-source nature of PDF readers makes LLM-driven specifications extraction extremely challenging.
Instead, we employ LLMs to analyze API manuals, generate API invocation rules, and infer intricate API relationships.

\vspace{0.02cm} \noindent \textbf{Grammar-based Fuzzing.}
Grammar-based fuzzing uses context-free grammars (CFGs) to generate syntactically valid inputs for structured data, targeting syntactic validity~\cite{DBLP:conf/sigsoft/grammarinator, DBLP:conf/icse/superion}, semantic validity~\cite{DBLP:conf/ndss/codealchemist, DBLP:conf/uss/langfuzz}, and behavioral exploration~\cite{DBLP:conf/ndss/fuzzili}.
Existing methods use different strategies: code fragment reuse (LangFuzz~\cite{DBLP:conf/uss/langfuzz}, CodeAlchemist~\cite{DBLP:conf/ndss/codealchemist}), custom IRs (DUMPLING~\cite{DBLP:conf/ndss/dumpling}), or direct CFGs~\cite{DBLP:conf/sigsoft/grammarinator, DBLP:conf/ndss/nautilus, DBLP:conf/issta/Amaya24}.
Mutation approaches include grammar-aware~\cite{DBLP:conf/ndss/nautilus, DBLP:conf/icse/superion, DBLP:conf/uss/langfuzz}, token-based~\cite{DBLP:conf/uss/tokenlevelfuzz, DBLP:conf/icse/superion}, and IR-guided mutations~\cite{DBLP:conf/ndss/fuzzili, DBLP:conf/uss/optfuzz}.
However, traditional grammar-based approaches embed only limited semantic constraints within grammar rules, failing to capture complex API relationships, including those defined in Section~\ref{sec:relation_definition}.
Our approach extends beyond syntactic generation by explicitly modeling these semantic constraints between API parameters and return values, then dynamically solving them to generate realistic, security-critical API usage sequences that traditional methods cannot achieve.

\section{Conclusion}

Fuzzing JavaScript engines of PDF readers faces critical challenges, including incomplete API specifications and limited API relationships inferred by previous works.
In this paper, we propose \ourtool, an LLM-assisted fuzzing tool to generate API function invocation rules based on documentation to improve fuzzing performance on the PDF reader's JavaScript engine. 
On three widely used PDF readers (Adobe Acrobat Reader, Foxit PDF Reader, and PDF-XChange Editor), we show that \ourtool improves fuzzing effectiveness over prior tools.
It achieves higher code coverage and uncovers more zero-day vulnerabilities.
Specifically, while current tools identify only 0--6 such vulnerabilities, \ourtool successfully uncovered 31 zero-day vulnerabilities.

\section*{Acknowledgment}
We would like to thank the anonymous reviewers for their constructive comments. 
This material is based in part upon work supported by DARPA under agreement N66001-22-2-4037. It is also supported by the National Science Foundation under grant no. 2229876 and is supported in part by funds provided by the U.S. National Science Foundation, by the Department of Homeland Security, and by IBM.
Yigitcan Kaya was supported by an appointment to the Intelligence Community Postdoctoral Research Fellowship Program at the National Institute of Standards and Technology administered by Oak Ridge Institute for Science and Education (ORISE) through an interagency agreement between the U.S. Department of Energy and the Office of the Director of National Intelligence (ODNI).
Any opinions, findings, and conclusions or recommendations expressed in this material are those of the author(s) and do not necessarily reflect the views of the DARPA, the U.S. National Science Foundation or its federal agency and industry partners.



\bibliographystyle{ACM-Reference-Format}
\bibliography{refs}

\appendix

\appendix

\section{Ethical Considerations}

\subsection{Identified Stakeholders and Impacts}

\smallskip \noindent \textbf{PDF Reader Vendors (Adobe, Foxit, PDF-XChange):}
Our research directly impacts these companies by identifying security vulnerabilities in their products. 
The discovery of 31 zero-day vulnerabilities could potentially affect their reputation and require resources for patching. 
However, our responsible disclosure approach provides them the opportunity to fix vulnerabilities before public disclosure, ultimately improving their product security.

\smallskip \noindent \textbf{End Users of PDF Readers: }
Users of the tested PDF readers are primary beneficiaries of this research, as the identified vulnerabilities (including those enabling arbitrary code execution) could be exploited by malicious actors to compromise user systems. 
Our research helps protect users by enabling vendors to patch these vulnerabilities before they can be widely exploited.

\subsection{Ethical Principles Analysis}

\smallskip \noindent \textbf{Beneficence:}
We applied consequentialist analysis weighing the benefits of vulnerability discovery against potential harms. 
The benefits include: 
(1) identification of serious security vulnerabilities that could prevent future exploitation, 
(2) methodological contributions to security research, and 
(3) improved security for millions of PDF reader users. 
These benefits substantially outweigh the limited harms associated with responsible disclosure.

\smallskip \noindent \textbf{Respect for Persons:}
 We respected the autonomy and interests of all stakeholders by: 
 (1) following responsible disclosure practices that respect vendors' rights to patch vulnerabilities before public disclosure, 
 (2) providing sufficient technical detail for reproduction and patching while avoiding exploitation guidance, and 
 (3) acknowledging the legitimate business interests of PDF reader vendors.

\subsection{Identified Harms and Mitigations}

\smallskip \noindent \textbf{Potential Harms:}
Premature or irresponsible disclosure could enable malicious exploitation of discovered vulnerabilities, potentially leading to system compromises, data breaches, or other security incidents.
Public disclosure of multiple vulnerabilities could negatively impact vendor reputations, even when disclosure is responsible.

\smallskip \noindent \textbf{Mitigations Implemented:}
We disclosed all 31 discovered vulnerabilities to the respective vendors immediately upon discovery, prior to any public disclosure or publication.
We worked with vendor security teams to establish appropriate disclosure timelines, allowing sufficient time for patch development and deployment.
We obtained confirmation from vendors regarding the validity of reported vulnerabilities and their patching status before publication.

\subsection{Decision Justification}

\smallskip \noindent \textbf{Decision to Proceed with Research:}
We determined that proceeding with this research was ethically justified based on several factors: 
(1) the critical importance of PDF reader security given their widespread deployment, 
(2) the demonstrated effectiveness of our methodology in discovering serious vulnerabilities, 
(3) our commitment to responsible disclosure practices, and 
(4) the substantial benefits to end-user security and the research community.

\smallskip \noindent \textbf{Decision to Publish:}
Publication is justified by: 
(1) the scientific value of our methodological contributions to LLM-assisted fuzzing, 
(2) responsible disclosure of all vulnerabilities prior to publication, 
(3) vendor acknowledgment and patching of critical vulnerabilities, 
(4) the educational value for the security research community, and 
(5) the transparency benefits for users and vendors regarding PDF reader security.

\section{Open Science}
Our artifact can be found at \url{https://github.com/ucsb-seclab/PDFuzzer}.

\end{document}